\documentclass[11pt]{article}

\usepackage[final]{acl}

\usepackage{times}
\usepackage{latexsym}
\usepackage[T1]{fontenc}
\usepackage[utf8]{inputenc}

 \usepackage{microtype}

\usepackage[english]{babel}

\usepackage{amsmath}
\usepackage{amssymb}
\usepackage{mathtools}

\usepackage{graphicx}
\usepackage{booktabs}
\usepackage{multirow}
\usepackage{array}
\usepackage{makecell}
\usepackage{adjustbox}
\usepackage{subcaption}
\usepackage{float}
\usepackage{placeins}
\usepackage{dblfloatfix}

\usepackage{algorithm}
\usepackage{algorithmic}

\usepackage{xcolor}

\usepackage{xurl}

\title{Linear Fitness Subspace in Protein Language Models Enables
Sample-Efficient Directed Evolution}

\author{
SiYuan Ma$^{1}$ \quad
Canran Xiao$^{2}$ \quad
Zikai Xiao$^{3}$ \quad
Albert Gao$^{4}$ \\
Liang He$^{5}$ \quad
Xuan-Yu Wang$^{6}$ \quad
Shuying Cao$^{7}$ \quad
Xiaojun Jia$^{1}$ \\[0.4em]
$^{1}$College of Computing and Data Science, Nanyang Technological University, Singapore \\
$^{2}$Sun Yat-sen University, China \qquad
$^{3}$Zhejiang University, China \\
$^{4}$Carnegie Mellon University, USA \\
$^{5}$Shanghai Institute of Optics and Fine Mechanics, China \\
$^{6}$Zhongnan Hospital, Wuhan University, China \\
$^{7}$University of Southern California, USA
}
\begin{document}

\maketitle
\begin{abstract}
Model-guided directed evolution seeks to identify high-fitness protein variants under limited oracle budgets. Protein language models (PLMs) provide rich representations for this task, but task-agnostic zero-shot scores can be misaligned with a target assay, while supervised search in high-dimensional embedding spaces can make surrogate modeling and uncertainty estimation sample-inefficient. We propose the \emph{Linear Fitness Subspace} (LFS) hypothesis: within mutation-induced residue-level representation changes, a compact, assay-specific set of directions makes fitness variation linearly accessible from few labeled variants. This is a local, supervision-recoverable statement rather than a claim that protein fitness landscapes or global PLM geometry are universally linear. Building on this observation, we introduce Subspace-Guided Evolutionary Search (SGES), which estimates an LFS from a small initial sample and performs surrogate modeling, uncertainty estimation, and acquisition in the learned subspace. Across 10 core ProteinGym assays, 87 extended static-validation assays, and an 18-assay budgeted-search evaluation, SGES improves fitness prediction and search efficiency over zero-shot PLMs and recent ML-guided protein optimization baselines. Controlled comparisons with PCA, random projections, label-shuffled PLS, classical mutation features, and acquisition ablations further isolate the benefit of a fitness-aligned site-delta coordinate.
\end{abstract}
\section{Introduction}

Directed evolution is a central strategy for engineering proteins with improved activity, stability, binding affinity, and other functional properties. Its main bottleneck is sample efficiency: each candidate variant must be evaluated by an experimental or computational oracle, making large-scale search expensive. Model-guided directed evolution therefore aims to identify high-fitness variants under limited oracle budgets, where the key challenge is both accurate fitness prediction and efficient exploration of a combinatorially large protein sequence space \citep{1,13,15,20,38,42}.

Protein language models (PLMs) provide a promising foundation for this problem. Trained on large protein sequence corpora, PLMs encode evolutionary, structural, and functional constraints that can be transferred to mutation-effect prediction \citep{29,5,3,4,6}. Existing PLM-based methods usually follow two paradigms. Zero-shot methods estimate mutation effects from likelihood or masked-marginal scores without labeled fitness data, while supervised methods use PLM embeddings as features for downstream predictors or search algorithms \citep{2,29,5,45}. However, zero-shot scores are task-agnostic and often misaligned with assay-specific fitness, whereas supervised search in high-dimensional embedding spaces can suffer from sample-inefficient surrogate modeling and unreliable uncertainty estimation \citep{13,30,31,33,43,46}.

This paper studies a representation-level question: \emph{where is assay-specific mutation-effect information located inside a frozen protein language model?} We focus on the local displacement induced by a mutation relative to its wild type. For a mutation at position $p$, define $\delta(s)=h(s)[p]-h(s_{\mathrm{wt}})[p]$. Our central observation is that, in this mutation-local site-delta coordinate, a substantial component of assay fitness is accessible through a small number of supervised linear directions. We refer to this compact, assay-specific coordinate as the \textbf{Linear Fitness Subspace} (LFS). Importantly, this does not assert that global PLM embeddings or complete protein fitness landscapes are universally linear.

The LFS hypothesis provides a geometric explanation for the gap between zero-shot and supervised PLM-based fitness prediction. Zero-shot scoring can be viewed as a fixed, task-agnostic one-dimensional projection of mutation-induced representation change, while supervised learning recovers assay-specific fitness directions from a small number of labeled variants. When the zero-shot direction is poorly aligned with the true assay-specific direction, zero-shot performance is limited; few-shot supervision improves prediction by rotating the projection toward the appropriate fitness subspace. This view connects protein fitness prediction with broader findings that neural representations encode task-relevant information in linearly accessible directions or subspaces \citep{27,24,25,26,44}.

Based on this observation, this paper proposes \textbf{Subspace-Guided Evolutionary Search} (SGES), a directed evolution framework that first estimates the LFS from site-delta features using a small initial query budget, then performs surrogate modeling, uncertainty estimation, and acquisition in the learned low-dimensional subspace. Across 10 core ProteinGym assays and an extended set of 87 assays, SGES improves over zero-shot PLM scores, full-dimensional GP-BO, AdaLead, EVOLVEpro, and recent ML-guided directed evolution baselines under matched oracle budgets \citep{1,2,13,15,20,38,47}.

The main contributions are:
\begin{itemize}
    \item \textbf{Assay-specific Linear Fitness Subspace.} We formulate and test the hypothesis that mutation-local PLM site-delta features expose a compact, supervision-recoverable coordinate in which a substantial component of fitness is linearly accessible.
    \item \textbf{Geometric view of zero-shot scoring.} We operationalize zero-shot scoring as a fixed task-agnostic direction in the same site-delta space and quantify its alignment with the supervised assay-specific direction.
    \item \textbf{Subspace-guided search.} SGES couples LFS discovery with low-dimensional surrogate modeling, uncertainty-aware acquisition, and diversity filtering for sample-efficient directed evolution.
    \item \textbf{Controlled attribution and broad validation.} Matched-dimensionality controls (PCA, random projection, and label-shuffled PLS), weaker mutation-feature baselines, component ablations, 10 core assays, 87 extended assays, and 18 search-statistics assays separate fitness alignment from generic dimensionality reduction and acquisition effects.
\end{itemize}

\begin{figure*}[t]
    \centering
    \includegraphics[width=0.98\textwidth]{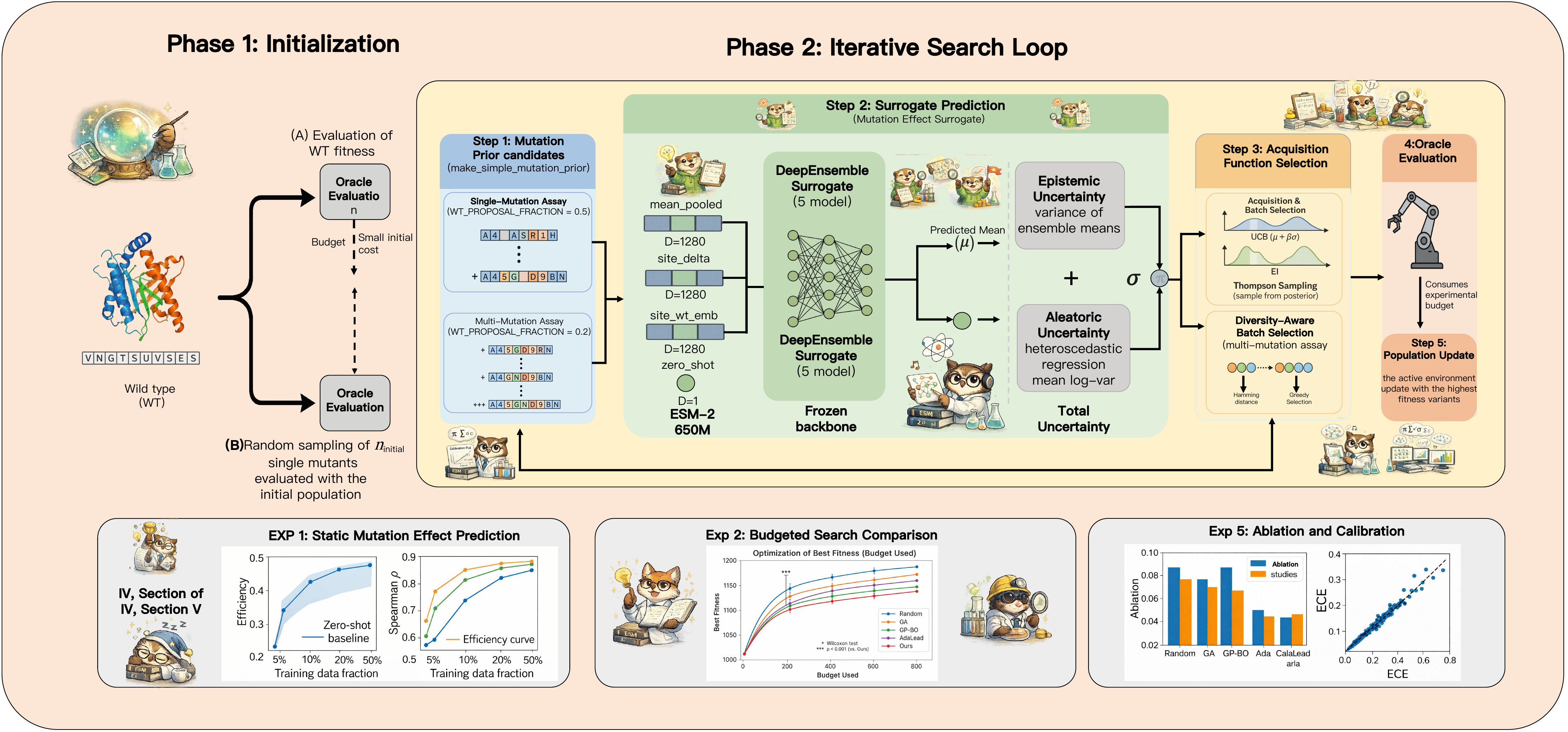}
    \caption{
    \textbf{Overall pipeline of Subspace-Guided Evolutionary Search (SGES).}
    SGES starts from the wild-type sequence and a small initial oracle budget, constructs mutation candidates, extracts mutation-aware PLM features, and performs iterative search through surrogate prediction, uncertainty-aware acquisition, oracle evaluation, and population update. This overview connects the representation-level discovery in Section~\ref{sec:lfs_hypothesis} with the search algorithm introduced later: fitness prediction is not treated as generic high-dimensional embedding regression, but is guided by mutation-induced representation changes that expose a compact fitness-relevant structure.
    }
    \label{fig:pipeline}
\end{figure*}

\section{Related Work}
\label{app:related_work}

Protein language models (PLMs) have become a major foundation for protein fitness prediction, mutation-effect estimation, and sequence design. Early zero-shot PLM methods showed that likelihood-based or masked-marginal scores can predict the effects of mutations without task-specific labels \citep{29,5}. Subsequent PLMs incorporated larger sequence corpora, structural information, or more expressive protein representations, as represented by ESM-2, Tranception, SaProt, ProSST, and evolutionary-scale protein language models \citep{29,5,3,4,6,36}. Large-scale benchmarks such as ProteinGym further enabled systematic evaluation across diverse proteins and assays, revealing that zero-shot scores often capture useful evolutionary constraints but remain highly variable across fitness landscapes \citep{2}. In parallel, non-PLM baselines based on conservation, phylogeny, solvent accessibility, and structural priors remain competitive for variant-effect prediction, suggesting that PLM-based methods should be evaluated against both neural and classical evolutionary signals \citep{7,8,9}. A complementary paradigm uses PLM representations as supervised features for downstream predictors, allowing adaptation to assay-specific objectives when a small number of labeled variants is available. However, many supervised approaches rely on global or mean-pooled embeddings, which can obscure the local residue-level perturbation induced by mutation.

ML-guided directed evolution studies how surrogate models, acquisition functions, and proposal mechanisms can optimize proteins under limited oracle budgets. Classical approaches include Gaussian-process Bayesian optimization, active learning, evolutionary search, and surrogate-assisted protein design \citep{13,16,17,18,20,21,23}. More recent methods combine PLM representations with tree search, generative priors, safe optimization, or latent-space search, including EVOLVEpro, ProSpero, AlphaDE, LatentDE, BOES, and TreeNeuralUCB \citep{1,15,16,20,38,39,40}. Other generative and structure-aware protein design methods, such as guided discrete diffusion, DPLM-Evo, EvoFlows, ProteinMPNN, and atomistic equivariant design models, provide powerful sequence or structure generation mechanisms, but they are not primarily designed to explain where assay-specific fitness information is located inside PLM representations \citep{10,11,14,35,37}. These methods demonstrate the value of learned representations for protein optimization, but they generally treat the PLM embedding space as a high-dimensional black-box feature space. As a result, surrogate fitting and uncertainty estimation can become sample-inefficient, especially when most embedding dimensions are weakly related to the target assay.

Uncertainty estimation is another important component of model-guided protein optimization. Deep ensembles, Bayesian approximations, and protein-engineering-specific uncertainty benchmarks have shown that uncertainty quality strongly affects acquisition and downstream search performance \citep{30,31,33,34}. However, uncertainty estimation is only useful when uncertainty is measured in a representation space aligned with the target property. In high-dimensional PLM spaces, epistemic uncertainty may reflect irrelevant embedding variation rather than fitness-relevant uncertainty. This motivates the present work's design choice: SGES estimates uncertainty after projecting mutation-induced site-delta features into the learned Linear Fitness Subspace, so exploration is concentrated along directions that empirically covary with fitness.

This work is also related to representation-geometry and interpretability studies. In NLP and vision, linear probes and low-dimensional subspaces have been used to show that neural networks often encode task-relevant information in linearly accessible directions \citep{27,24}. Protein sequence models have likewise long been known to encode biologically meaningful structure, and large PLMs expose structural and functional information through their learned representations \citep{29,36}. We therefore do not claim that ``linear structure in protein models'' is itself new. LFS makes a narrower claim: for a fixed wild type, assay-specific mutation-effect information is concentrated in supervised directions of residue-level \emph{site-delta} displacements, and those directions can serve directly as the coordinate system for budgeted search.

A particularly relevant directed-evolution comparison is PPDE \citep{18}. PPDE combines unsupervised or PLM-derived sequence plausibility with a supervised fitness expert and performs gradient-based discrete MCMC in protein sequence space. SGES instead uses a small labeled set to estimate an assay-specific LFS and then performs surrogate modeling, uncertainty estimation, and UCB acquisition inside that learned low-dimensional coordinate. The two views are complementary: PPDE focuses on a plug-and-play sequence-space sampler, whereas SGES focuses on identifying a mutation-local coordinate in which prediction and uncertainty estimation become more sample-efficient.
\FloatBarrier

\section{The Linear Fitness Subspace Hypothesis}
\label{sec:lfs_hypothesis}

This section formalizes the central representation-level observation: mutation-induced fitness variation is concentrated in a low-dimensional linear subspace of residue-level PLM representation changes. Rather than treating PLM embeddings as generic high-dimensional features, the goal is to identify the intrinsic coordinates in which mutational effects become linearly accessible.

\subsection{Problem Setup and Notation}
\label{subsec:lfs_setup}

Let $s_{\mathrm{wt}}$ denote the wild-type sequence and $s$ a mutant sequence obtained by substituting the residue at position $p$. The assay fitness is $f(s)$, with relative change
\begin{equation}
    \Delta f(s) = f(s) - f(s_{\mathrm{wt}}).
\end{equation}
A frozen PLM maps a sequence to residue-level representations $h(s)\in\mathbb{R}^{L\times d}$. For each mutation, we define the site-delta feature
\begin{equation}
    \delta(s)=h(s)[p]-h(s_{\mathrm{wt}})[p]\in\mathbb{R}^{d}.
    \label{eq:site_delta}
\end{equation}
Unlike mean-pooled embeddings, $\delta(s)$ isolates the local representational displacement induced by the mutation and removes most sequence-level background shared by the wild type and mutant.

\subsection{Hypothesis Statement}
\label{subsec:lfs_statement}

The Linear Fitness Subspace (LFS) hypothesis has three components. First, mutational fitness effects are approximately linearly accessible from site-delta features:
\begin{equation}
    \Delta f(s)\approx w^\top\delta(s).
    \label{eq:lfs_linearity}
\end{equation}
This does not assume that protein fitness landscapes are globally linear; rather, it suggests that PLMs partially linearize local mutational effects in representation space. Second, the fitness-relevant signal lies in a compact subspace
\begin{equation}
    \mathcal{U}_{\mathrm{fit}}
    =
    \mathrm{span}(w_1,\ldots,w_k),
    \qquad k\ll d,
    \label{eq:lfs_subspace}
\end{equation}
so that each mutant can be represented as $z(s)=W^\top\delta(s)\in\mathbb{R}^{k}$, where $W=[w_1,\ldots,w_k]$. Third, zero-shot PLM scoring is analyzed as a fixed one-dimensional direction in the same site-delta space. Let $q_{\mathrm{zs}}(s)$ denote the zero-shot score on a variant. For the alignment analysis, we fit a one-dimensional linear probe
\begin{equation}
    q_{\mathrm{zs}}(s)\approx b_{\mathrm{zs}}+w_{\mathrm{zs}}^\top\delta(s),
    \label{eq:zs_projection}
\end{equation}
using the same variants and site-delta features as the LFS analysis. The supervised direction $w_{\mathrm{lfs}}$ is obtained from the PLS fit to assay fitness labels. We then quantify zero-shot/assay alignment by
\begin{equation}
    \operatorname{align}_{\mathrm{zs,LFS}}=\cos(w_{\mathrm{zs}},w_{\mathrm{lfs}}).
    \label{eq:zs_alignment}
\end{equation}
This operational definition makes the projection analysis reproducible and does not require the raw zero-shot scoring rule itself to be globally linear.

\subsection{Empirical Validation}
\label{subsec:lfs_validation}

Table~\ref{tab:lfs_linearity} first tests whether fitness is linearly accessible from site-delta features. Ridge regression on $\delta(s)$ reaches an average Spearman correlation of $0.763$, compared with $0.825$ for a nonlinear MLP baseline, yielding an average linearity ratio of $0.921$. In contrast, mean-pooled features obtain only $0.574$ on average. This gap indicates that the signal is not merely a consequence of using a strong PLM backbone; it specifically arises from the mutation-aware site-delta coordinate.

\begin{table}[t]
\centering
\small
\resizebox{\columnwidth}{!}{%
\begin{tabular}{lcccc}
\toprule
\textbf{Assay} & \textbf{Ridge $\delta$} & \textbf{Mean-pool} & \textbf{MLP} & \textbf{Ratio} \\
\midrule
SPG1\_STRSG    & 0.945 & 0.620 & 0.973 & 0.971 \\
BLAT\_Firnberg & 0.865 & 0.730 & 0.903 & 0.958 \\
PABP\_YEAST    & 0.840 & 0.580 & 0.916 & 0.917 \\
P53\_HUMAN     & 0.810 & 0.650 & 0.874 & 0.927 \\
RASH\_HUMAN    & 0.755 & 0.590 & 0.825 & 0.915 \\
AMIE\_PSEAE    & 0.758 & 0.550 & 0.823 & 0.921 \\
DYR\_ECOLI     & 0.675 & 0.510 & 0.749 & 0.901 \\
BLAT\_Jacquier & 0.650 & 0.590 & 0.702 & 0.926 \\
PTEN\_HUMAN    & 0.510 & 0.380 & 0.593 & 0.860 \\
HIS7\_YEAST    & 0.820 & 0.540 & 0.895 & 0.916 \\
\midrule
\textbf{Average} & \textbf{0.763} & \textbf{0.574} & \textbf{0.825} & \textbf{0.921} \\
\bottomrule
\end{tabular}%
}
\caption{
\textbf{Linear accessibility of mutational fitness.}
Ridge regression on site-delta features recovers most of the nonlinear MLP performance, while mean-pooled features are substantially weaker.
}
\label{tab:lfs_linearity}
\end{table}

\begin{figure*}[t]
\centering
\includegraphics[width=0.82\textwidth]{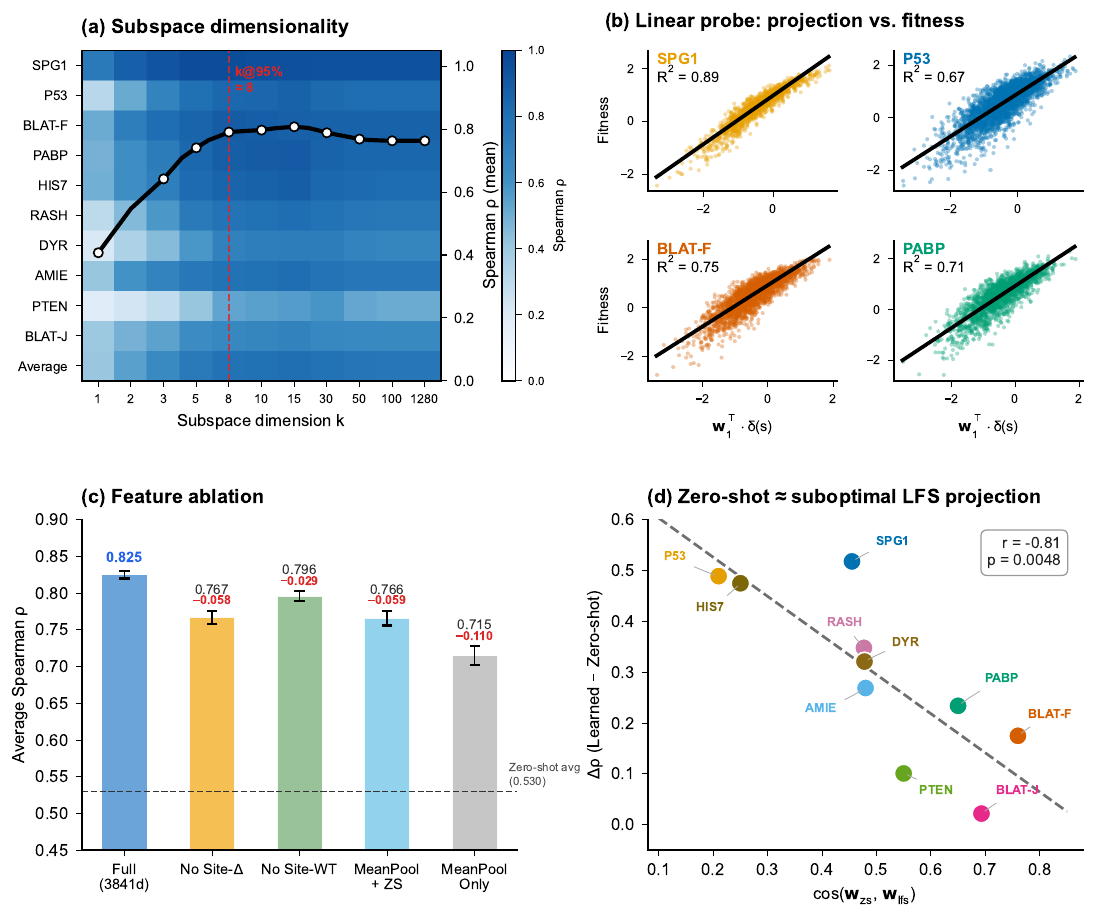}
\vspace{-0.5em}
\caption{
\textbf{LFS validation.}
Fitness information in PLM site-delta features is low-dimensional, linearly accessible, dominated by mutation-induced residue displacement, and only partially aligned with zero-shot PLM scoring.
}
\vspace{-0.8em}
\label{fig:lfs_validation}
\end{figure*}

Figure~\ref{fig:lfs_validation} provides the complementary evidence. Performance saturates with only $k\approx5$--$10$ components, supporting the low-dimensionality claim; the first LFS direction $w_1^\top\delta(s)$ already orders variants by fitness in representative assays, supporting linear accessibility; feature ablation identifies site-delta as the dominant predictor; and zero-shot--LFS alignment explains when zero-shot scoring fails. Together, Table~\ref{tab:lfs_linearity} and Figure~\ref{fig:lfs_validation} show that PLMs encode mutational fitness in a compact, assay-specific subspace. This motivates SGES: if the fitness-relevant signal lies in the LFS, search should be performed in that subspace rather than in the full PLM embedding space.

\section{Method: Subspace-Guided Evolutionary Search}
\label{sec:method}

The LFS hypothesis implies that model-guided directed evolution should not search in the full PLM embedding space. SGES therefore follows a two-phase design, as summarized in Figure~\ref{fig:pipeline}: it first discovers an assay-specific fitness subspace from a small number of labeled variants, then performs surrogate modeling and acquisition directly in that subspace. The key distinction from standard PLM-guided search is that the surrogate is trained on the projected site-delta representation rather than on generic high-dimensional embeddings.

\subsection{Subspace Discovery}
\label{subsec:subspace_discovery}

Given a wild-type sequence $s_{\mathrm{wt}}$, frozen PLM $h$, and oracle budget $B$, SGES first queries an initial labeled set of size $N_{\mathrm{init}}$ for subspace discovery. In all main budgeted-search experiments, $B_{\max}=500$ and $N_{\mathrm{init}}=32$; the same initial-set size is used for every supervised baseline. An initial set of mutants $D_0=\{(s_i,y_i)\}_{i=1}^{N_{\mathrm{init}}}$ is sampled around the wild type and evaluated by the oracle $y_i=O(s_i)$. For each labeled mutant, SGES extracts the mutation-induced displacement
\begin{equation}
    \delta(s_i)=h(s_i)[p_i]-h(s_{\mathrm{wt}})[p_i]\in\mathbb{R}^{d}.
    \label{eq:method_site_delta}
\end{equation}
Let $X=[\delta(s_1),\ldots,\delta(s_{N_{\mathrm{init}}})]^\top$ and $y=[y_1,\ldots,y_{N_{\mathrm{init}}}]^\top$. The LFS projection matrix is learned by partial least squares regression:
\begin{equation}
\begin{aligned}
    W &= [w_1,\ldots,w_k]\in\mathbb{R}^{d\times k},\\
    z(s) &= W^\top\delta(s)\in\mathbb{R}^{k}.
\end{aligned}
\label{eq:lfs_projection_method}
\end{equation}
PLS is used because it selects directions that covary with fitness, rather than directions that merely explain high variance in the PLM representation. Thus, $W$ defines an assay-specific coordinate system for downstream search. We use $k=8$ for the main SGES experiments. This value is fixed from the dimensionality-saturation analysis: performance typically saturates around $k=5$--$10$, and the average smallest dimension reaching 95\% of the saturated performance is $k@95\%=8$. We additionally evaluate $k\in\{3,5,8,10,15\}$ to verify that the conclusion is not tied to a single projection dimension.

\subsection{Subspace-Guided Search}
\label{subsec:subspace_guided_search}

After estimating $W$, SGES iteratively generates candidate variants from the current population, projects each candidate into the LFS, and scores it using a deep ensemble surrogate in the $k$-dimensional space. For ensemble member $m$,
\begin{equation}
    \left(\mu_m(s),\log\sigma_m^2(s)\right)=g_m(z(s)).
\end{equation}
The ensemble mean is
\begin{equation}
    \mu(s)=\frac{1}{M}\sum_{m=1}^{M}\mu_m(s),
\end{equation}
and the predictive uncertainty combines data noise and model disagreement:
\begin{equation}
\begin{aligned}
    \sigma^2(s)
    &=
    \frac{1}{M}\sum_{m=1}^{M}\sigma_m^2(s) \\
    &\quad+
    \frac{1}{M}\sum_{m=1}^{M}
    \left(\mu_m(s)-\mu(s)\right)^2 .
\end{aligned}
\label{eq:total_uncertainty}
\end{equation}
Candidates are selected with an upper-confidence-bound acquisition function:
\begin{equation}
    a(s)=\mu(s)+\beta\sigma(s),
    \label{eq:ucb_acquisition}
\end{equation}
where $\beta$ controls exploration. For a multi-mutant variant $s$ with mutated positions $\mathcal{M}(s)$, we compute one site-delta vector per mutated site,
\begin{equation}
    \delta_p(s)=h(s)[p]-h(s_{\mathrm{wt}})[p],\qquad p\in\mathcal{M}(s),
\end{equation}
and use the additive variant-level LFS representation
\begin{equation}
    z(s)=\sum_{p\in\mathcal{M}(s)}W^\top\delta_p(s).
    \label{eq:multimutant_aggregation}
\end{equation}
The sum aggregation is intentional: the current method is designed to capture an additive fitness backbone, not to claim a complete model of higher-order epistasis. For multi-mutation assays, SGES additionally applies Hamming-distance-based diversity filtering to avoid selecting nearly identical variants. Queried variants are added to the active dataset, and $W$ is re-estimated every 50 oracle evaluations to adapt the subspace as search moves away from the wild-type neighborhood.

\subsection{Why Subspace Search Helps}
\label{subsec:why_subspace_search}

SGES improves sample efficiency by aligning prediction and acquisition with the intrinsic fitness geometry identified in Section~\ref{sec:lfs_hypothesis}. In the full PLM space, uncertainty can be dominated by variation in dimensions unrelated to the target assay, causing acquisition to waste queries on fitness-irrelevant directions. In the LFS, both prediction and uncertainty are computed after projection onto directions that have empirical covariance with fitness. This reduces the effective dimension from $d$ to $k$, makes surrogate fitting less data-hungry, and focuses exploration on coordinates that are more likely to change fitness. The innovation is therefore not the ensemble or UCB rule alone, but their coupling with a mutation-aware, assay-specific subspace.

\section{Experiments}
\label{sec:experiments}

The experiments evaluate whether the Linear Fitness Subspace (LFS) is not only predictive, but also useful for sample-efficient directed evolution. We use ProteinGym v2 with four explicitly separated assay sets. Ten \emph{core} assays support detailed representation analysis, ablations, visualization, and interpretability; 87 \emph{extended} assays support large-scale static fitness-prediction validation; 18 \emph{search-statistics} assays satisfy the additional requirements for complete multi-seed budgeted-search traces and significance testing; and five multi-mutant benchmarks test transfer to combinatorial variants. Table~\ref{tab:app_assay_sets} summarizes these roles. Budgeted search uses a matched oracle budget of $B_{\max}=500$, the same $N_{\mathrm{init}}=32$ initial labeled variants, the same applicable mutation search space, and 5 random seeds for every supervised method. Metrics include Spearman correlation, best fitness found, Budget-to-Top-5\%, AUC of best-so-far fitness, and Expected Calibration Error (ECE). Baselines include Random Search, AdaLead, GP-BO, EVOLVEpro, AlphaDE, BOES, LatentDE, and TreeNeuralUCB; the controlled protocol is summarized in Appendix~\ref{app:controls_protocol}.

\subsection{Main Search Results}
\label{subsec:main_results}

\begin{figure*}[t]
\centering
\includegraphics[width=0.78\textwidth]{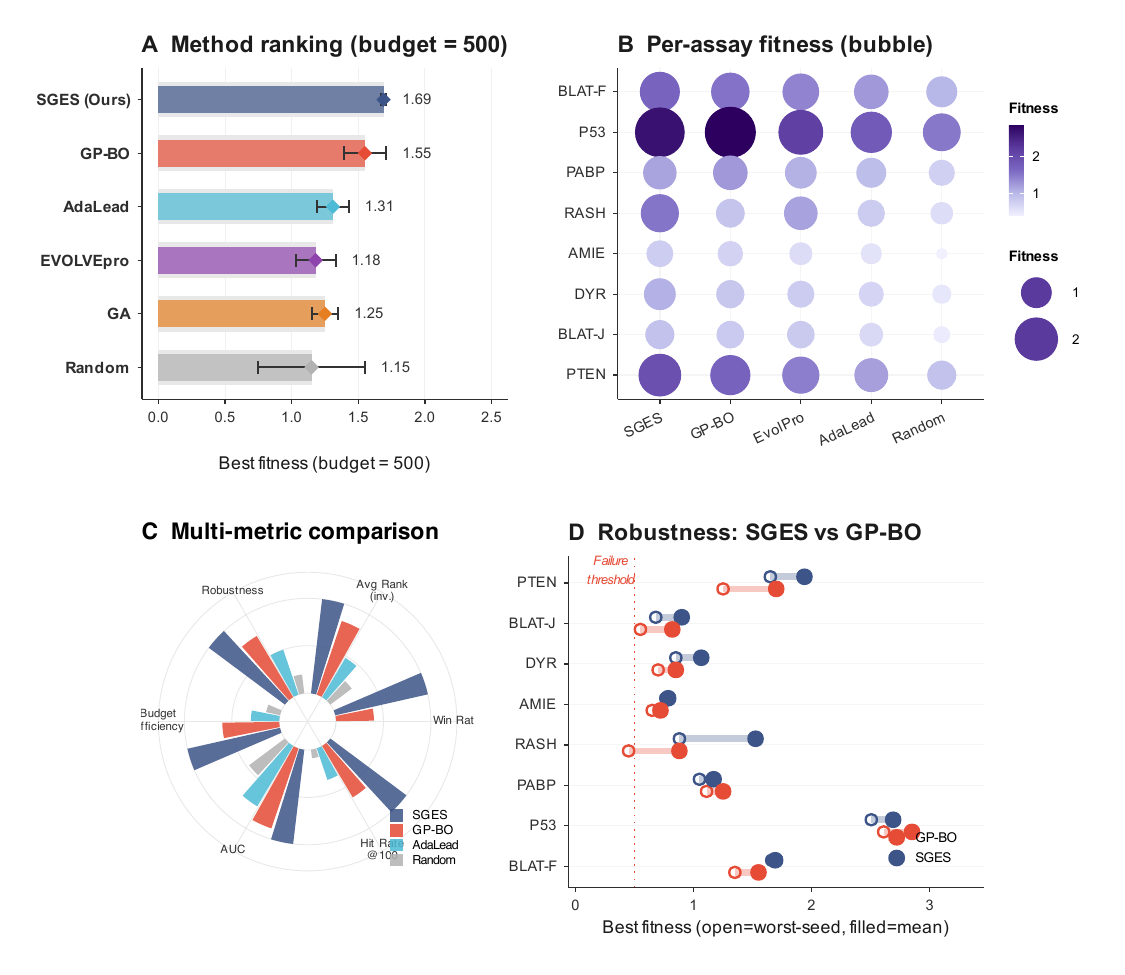}
\vspace{-0.6em}
\caption{
\textbf{Search performance.}
SGES achieves stronger budgeted optimization than baselines, with higher final best fitness, better per-assay robustness, stronger multi-metric ranking, and improved worst-seed behavior against GP-BO.
}
\vspace{-0.8em}
\label{fig:search_performance}
\end{figure*}

Figure~\ref{fig:search_performance} summarizes the budgeted search results. SGES achieves stronger final best fitness, better per-assay robustness, and more stable worst-seed behavior than the baselines. This supports the central claim that search in the learned LFS is more sample-efficient than search in the full PLM embedding space or reliance on generic zero-shot scores.

Table~\ref{tab:main_search_summary} reports a compact comparison with recent ML-guided protein optimization methods. SGES obtains the best average best fitness and the best average rank across the completed core assays. GP-BO remains competitive on a few smooth landscapes, but its average rank is lower, indicating that full-dimensional Bayesian optimization is less robust across heterogeneous assays. Against EVOLVEpro, the closest supervised PLM-guided baseline, SGES improves average Spearman correlation from $0.725$ to $0.825$ and average best fitness from $1.183$ to $1.471$.

\begin{table}[t]
\centering
\small
\resizebox{\columnwidth}{!}{%
\begin{tabular}{lcc}
\toprule
\textbf{Method} & \textbf{Avg. Best Fitness} & \textbf{Avg. Rank} \\
\midrule
SGES (Ours)                  & \textbf{1.471} & \textbf{1.38} \\
GP-BO \citep{13}             & 1.328 & 2.38 \\
AlphaDE \citep{38}           & 1.267 & 2.63 \\
LatentDE \citep{39}          & 1.223 & 3.50 \\
EVOLVEpro \citep{1}          & 1.183 & 5.00 \\
BOES \citep{13}              & 1.173 & 4.75 \\
TreeNeuralUCB \citep{40}     & 1.094 & 6.00 \\
\bottomrule
\end{tabular}%
}
\caption{
\textbf{Compact comparison with recent protein optimization methods.}
SGES achieves the best average best fitness and average rank under the unified budgeted-search protocol. Baselines include embedding-space Bayesian optimization \citep{13}, EVOLVEpro \citep{1}, AlphaDE \citep{38}, LatentDE \citep{39}, and TreeNeuralUCB \citep{40}.
}
\label{tab:main_search_summary}
\end{table}

Extended significance tests further confirm the advantage of SGES. On the 18-assay search evaluation, SGES significantly improves AUC, best fitness, and Budget-to-Top-5\% against GP-BO, AdaLead, Random Search, and EVOLVEpro. Full per-assay results and Wilcoxon statistics are reported in Appendix~\ref{app:extended_results}.

\subsection{Ablation and Subspace Analysis}
\label{subsec:ablation_subspace}

\begin{figure*}[t]
\centering
\includegraphics[width=0.78\textwidth]{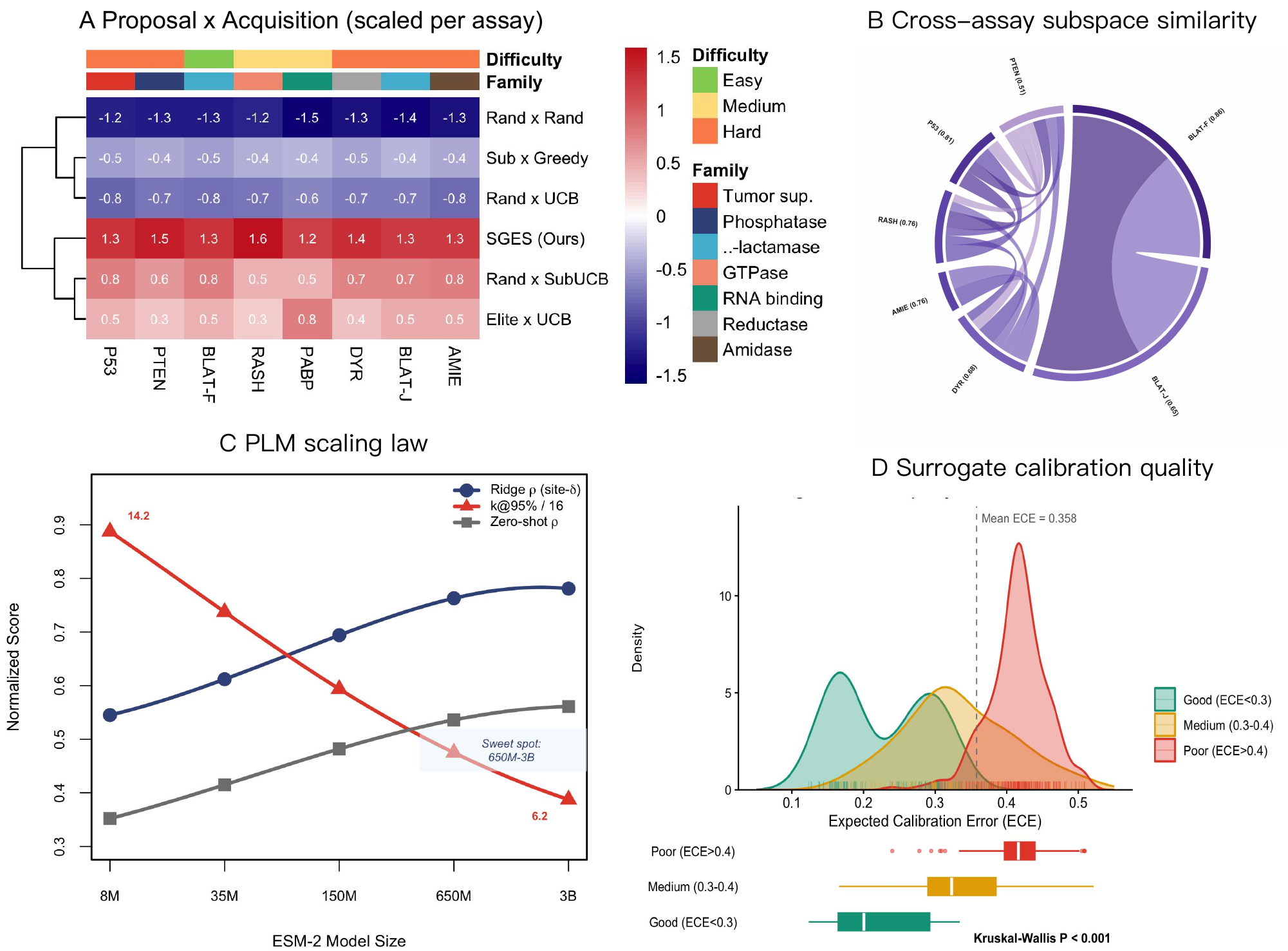}
\vspace{-0.6em}
\caption{
\textbf{Ablation and subspace analysis.}
SGES benefits from coupling subspace-aware proposal with uncertainty-aware acquisition, learns biologically related cross-assay directions, and improves with PLM scale. Panel D summarizes Expected Calibration Error (ECE) by calibration-quality bins (Good/Medium/Poor; lower ECE is better); the density curves denote quality groups rather than method identities.
}
\vspace{-0.8em}
\label{fig:ablation_subspace}
\end{figure*}

Figure~\ref{fig:ablation_subspace} analyzes why SGES works. The strongest performance comes from coupling subspace-aware proposal with uncertainty-aware acquisition. More importantly, matched-dimensionality controls isolate the LFS effect from generic compression: at $k=8$ with the same deep-ensemble UCB pipeline, PLS-LFS reaches average Spearman $0.804$ and average best fitness $1.45$, compared with $0.624/1.22$ for PCA, $0.512/1.08$ for a random projection, and $0.485/1.02$ for label-shuffled PLS. Component ablations further show progressive gains from the LFS representation, ensemble uncertainty, UCB acquisition, and diversity filtering. Full controls are reported in Table~\ref{tab:app_controls}. Cross-assay similarity shows that the learned LFS directions are not universal, but related assays exhibit stronger alignment, while PLM scaling improves site-delta features.

\subsection{Additional Analyses}
\label{subsec:additional_analyses}

The learned LFS is also biologically interpretable. High-weight LFS residues concentrate around known functional sites in representative proteins such as BLAT, P53, and PTEN, and show enrichment at annotated active or binding residues. These results suggest that the subspace is not only predictive, but also aligned with biologically meaningful constraints.

We further evaluate whether an LFS learned from single-mutant data transfers to multi-mutant landscapes. Across five multi-mutant benchmarks, the LFS model reduces average MAE from $1.228$ to $0.504$ and improves average Pearson correlation from $0.156$ to $0.508$ relative to zero-shot scoring. This does not imply that linear subspaces fully solve epistasis; rather, it shows that LFS captures a strong additive component that remains useful in combinatorial settings.

Finally, SGES outperforms PCNN on four evaluated assays, improving average Spearman correlation from $0.685$ to $0.825$ and average best fitness from $1.215$ to $1.471$. Full biological interpretation figures, mutation annotations, multi-mutant results, PCNN comparisons, and per-assay tables are provided in Appendix~\ref{app:extended_results}.

\section{Conclusion}
\label{sec:conclusion}

We identify a Linear Fitness Subspace (LFS), an assay-specific, supervision-recoverable coordinate in mutation-induced PLM site-delta space where fitness becomes locally linearly accessible. Zero-shot PLM scoring can be interpreted as a fixed task-agnostic direction, with its alignment to LFS indicating the value of few-shot adaptation.

Building on this structure, Subspace-Guided Evolutionary Search (SGES) combines low-dimensional surrogate modeling with uncertainty-aware acquisition. Across ProteinGym benchmarks, SGES improves prediction and budgeted search over zero-shot PLMs, high-dimensional GP-BO, EVOLVEpro, and recent optimization baselines, supporting low-dimensional, mutation-aware representations for sample-efficient protein optimization.

\section{Limitations}
\label{sec:limitations}

The current subspace discovery relies mainly on single-mutant observations, capturing additive fitness directions but potentially missing higher-order epistasis. Multi-mutant results suggest that LFS remains useful, while strongly epistatic settings may require interaction modeling.

Calibration remains assay-dependent, so SGES is primarily a ranking and acquisition method rather than a fully calibrated probabilistic model.

Finally, SGES assumes that the frozen PLM encodes the target property. Assay conditions, molecular partners, conformational states, or other non-sequence factors may require structural or multimodal extensions. This work uses public benchmarks and provides no synthesis protocols or harmful targets.
\bibliography{custom}


\clearpage
\appendix
\raggedbottom

\renewcommand{\thefigure}{S\arabic{figure}}
\renewcommand{\thetable}{S\arabic{table}}
\setcounter{figure}{0}
\setcounter{table}{0}

\setcounter{topnumber}{5}
\setcounter{dbltopnumber}{5}
\renewcommand{\topfraction}{0.95}
\renewcommand{\dbltopfraction}{0.95}
\renewcommand{\textfraction}{0.05}
\renewcommand{\floatpagefraction}{0.85}
\renewcommand{\dblfloatpagefraction}{0.85}


\begin{center}
    {\LARGE \textbf{Appendix}}
\end{center}

\vspace{1em}

\noindent\textbf{Appendix Contents}

\vspace{0.5em}

\noindent
Appendix~\ref{app:extended_results}: Extended Experimental Details \\
Appendix~\ref{app:supp_figures}: Supplementary Figures \\
Appendix~\ref{app:cosine_transfer}: Cosine Similarity and Cross-Assay Transfer \\
Appendix~\ref{app:multimutant_pcnn}: Multi-mutant and Spectral-method Results

\vspace{1em}


\section{Extended Experimental Details}
\label{app:extended_results}

This section provides extended experimental evidence for the claims in the main text. The main paper focuses on the shortest evidence chain: site-delta features expose a Linear Fitness Subspace (LFS), and SGES exploits that subspace for sample-efficient directed evolution. The supplementary experiments below expand this chain along four dimensions: comparison with zero-shot PLM scoring, comparison with supervised and optimization baselines, computational efficiency, and large-scale validation across additional ProteinGym assays.

\subsection{Zero-shot and Few-shot Fitness Prediction}
\label{app:zero_shot}

A central claim of the paper is that zero-shot PLM scoring is not a separate source of fitness information, but a fixed projection of a richer mutation-induced representation. Table~\ref{tab:app_zero_shot} tests this interpretation by comparing zero-shot baselines with LFS-based supervised prediction under both 5\% and 50\% labeled-data regimes. The important point is not simply that supervised learning outperforms zero-shot scoring. Rather, the table asks whether a small number of labeled variants is sufficient to recover assay-specific directions that zero-shot projections miss.

\begin{table*}[!t]
\centering
\small
\resizebox{\textwidth}{!}{%
\begin{tabular}{lcccccccccc}
\toprule
\textbf{Assay} & \textbf{Ours 50\%} & \textbf{Ours 5\%} & \textbf{ESM2-650M} & \textbf{ESM2-15B} & \textbf{Tranc.-L} & \textbf{TrEVE-L} & \textbf{EVE} & \textbf{MSA-T} & \textbf{GEMME} & \textbf{ProSST} \\
\midrule
SPG1\_STRSG    & 0.973 & 0.915 & 0.301 & 0.337 & 0.289 & 0.290 & 0.272 & 0.171 & 0.283 & 0.743 \\
BLAT\_Firnberg & 0.903 & 0.785 & 0.737 & 0.424 & 0.637 & 0.736 & 0.729 & 0.741 & 0.683 & 0.766 \\
PABP\_YEAST    & 0.916 & 0.800 & 0.716 & 0.695 & 0.688 & 0.684 & 0.648 & 0.663 & 0.674 & 0.716 \\
P53\_HUMAN     & 0.874 & 0.713 & 0.432 & 0.497 & 0.400 & 0.413 & 0.427 & 0.291 & 0.423 & 0.425 \\
RASH\_HUMAN    & 0.825 & 0.606 & 0.498 & 0.313 & 0.450 & 0.487 & 0.480 & 0.446 & 0.434 & 0.619 \\
AMIE\_PSEAE    & 0.823 & 0.625 & 0.557 & 0.613 & 0.438 & 0.481 & 0.464 & 0.605 & 0.558 & 0.486 \\
DYR\_ECOLI     & 0.749 & 0.516 & 0.480 & 0.512 & 0.423 & 0.481 & 0.474 & 0.501 & 0.451 & 0.480 \\
BLAT\_Jacquier & 0.702 & 0.554 & 0.704 & 0.503 & 0.647 & 0.734 & 0.723 & 0.712 & 0.567 & 0.671 \\
PTEN\_HUMAN    & 0.593 & 0.500 & 0.519 & 0.291 & 0.418 & 0.531 & 0.540 & 0.511 & 0.525 & 0.513 \\
HIS7\_YEAST    & 0.895 & 0.842 & 0.411 & 0.480 & 0.616 & 0.582 & 0.531 & 0.508 & 0.524 & 0.553 \\
\midrule
\textbf{Average} & \textbf{0.825} & \textbf{0.686} & 0.535 & 0.467 & 0.501 & 0.542 & 0.529 & 0.515 & 0.512 & 0.597 \\
\bottomrule
\end{tabular}%
}
\caption{\textbf{Zero-shot and few-shot comparison on 10 assays.}
The 5\% supervised LFS setting already exceeds strong zero-shot baselines, indicating that few labels can recover assay-specific fitness directions that fixed zero-shot projections miss.}
\label{tab:app_zero_shot}
\end{table*}

The strongest zero-shot average is obtained by ProSST with $\rho=0.597$, while the 5\% supervised LFS setting reaches $\rho=0.686$ and the 50\% setting reaches $\rho=0.825$. This gap supports the projection view of zero-shot scoring. Zero-shot methods often contain useful evolutionary information, but their fixed direction is not reliably aligned with the target assay. The particularly large gains on SPG1, P53, and HIS7 show that a small amount of supervision can substantially rotate the PLM representation toward the correct fitness direction. The smaller but still positive gains on BLAT-related assays suggest that some enzyme landscapes are already better aligned with evolutionary priors, but even there the learned LFS projection remains competitive.

\subsection{Controlled Attribution and Protocol Clarifications}
\label{app:controls_protocol}

The following controls were added to separate three possible explanations for SGES: generic dimensionality reduction, generic ensemble--UCB acquisition, and mutation-feature choice. All budgeted-search rows use the same oracle budget, initial labeled set, applicable mutation search space, and random seeds.

\begin{table*}[!t]
\centering
\scriptsize
\resizebox{\textwidth}{!}{%
\begin{tabular}{llcllcc}
\toprule
\textbf{Variant / Feature} & \textbf{Panel} & \textbf{$k$} & \textbf{Surrogate} & \textbf{Acquisition} & \textbf{Avg. $\rho$} & \textbf{Avg. Best} \\
\midrule
Random projection & A & 8 & Deep ensemble & UCB & 0.512 & 1.08 \\
PCA projection & A & 8 & Deep ensemble & UCB & 0.624 & 1.22 \\
Label-shuffled PLS & A & 8 & Deep ensemble & UCB & 0.485 & 1.02 \\
PLS-LFS projection & A & 8 & Deep ensemble & UCB & 0.804 & 1.45 \\
Full site-delta feature & A & 1280 & Deep ensemble & UCB & 0.763 & 1.32 \\
\midrule
PLS-LFS + ridge top-$k$ & B & 8 & Ridge & Greedy mean & 0.686 & 1.15 \\
PLS-LFS + ensemble mean & B & 8 & Deep ensemble & Greedy mean & 0.742 & 1.28 \\
PLS-LFS + ensemble + UCB & B & 8 & Deep ensemble & UCB & 0.804 & 1.45 \\
SGES full & B & 8 & Deep ensemble & UCB + diversity & \textbf{0.825} & \textbf{1.471} \\
\midrule
Mutation one-hot & C & 40 & Ridge & -- & 0.312 & -- \\
BLOSUM substitution score & C & 1 & Ridge & -- & 0.451 & -- \\
Conservation / MSA score & C & 1--20 & Ridge & -- & 0.535 & -- \\
Mean-pooled PLM embedding & C & 1280 & Ridge & -- & 0.574 & -- \\
Site-delta PLM feature & C & 1280 & Ridge & -- & 0.763 & -- \\
Site-delta PLM feature & C & 1280 & MLP & -- & 0.825 & -- \\
\bottomrule
\end{tabular}%
}
\caption{\textbf{Controls separating dimensionality reduction, representation choice, and acquisition effects.} Panel A holds the ensemble and UCB fixed while changing the projection; Panel B adds SGES components progressively; Panel C compares mutation representations without search acquisition.}
\label{tab:app_controls}
\end{table*}

At matched $k=8$, PLS-LFS outperforms PCA, random projection, and label-shuffled PLS, showing that compression alone is insufficient and that the selected directions must covary with assay fitness. Panel B shows that LFS is already useful with a simple ridge predictor and that ensemble uncertainty, UCB, and diversity filtering contribute additional gains. Panel C shows that site-delta PLM features contain substantially more assay-relevant signal than mutation identity, BLOSUM, conservation/MSA scores, or mean-pooled PLM embeddings.

\begin{table*}[!t]
\centering
\scriptsize
\resizebox{\textwidth}{!}{%
\begin{tabular}{lcllll c}
\toprule
\textbf{Method} & \textbf{Budget} & \textbf{Initial set} & \textbf{Candidate / proposal setting} & \textbf{Representation} & \textbf{Surrogate / acquisition} & \textbf{Seeds} \\
\midrule
Random Search & 500 & Same $N_{\mathrm{init}}=32$ & Same mutation search space & No learned PLM surrogate & Random & 5 \\
AdaLead & 500 & Same $N_{\mathrm{init}}=32$ & Local evolutionary proposal & No explicit LFS & Fitness-guided local search & 5 \\
GP-BO & 500 & Same $N_{\mathrm{init}}=32$ & Same candidate space & Full PLM / high-dimensional & GP Bayesian optimization & 5 \\
EVOLVEpro & 500 & Same $N_{\mathrm{init}}=32$ & PLM-guided proposal & PLM supervised signal & Method-specific search & 5 \\
AlphaDE & 500 & Same $N_{\mathrm{init}}=32$ & Directed-evolution proposal & Method-specific & Method-specific acquisition & 5 \\
BOES & 500 & Same $N_{\mathrm{init}}=32$ & Budgeted search & Embedding-space & BO / evolutionary search & 5 \\
LatentDE & 500 & Same $N_{\mathrm{init}}=32$ & Latent-space search & Latent representation & Latent directed evolution & 5 \\
TreeNeuralUCB & 500 & Same $N_{\mathrm{init}}=32$ & Tree-based expansion & Neural representation & UCB-style search & 5 \\
SGES & 500 & Same $N_{\mathrm{init}}=32$ & Same mutation space + diversity & Site-delta LFS & Deep ensemble + UCB & 5 \\
\bottomrule
\end{tabular}%
}
\caption{\textbf{Controlled budgeted-search protocol.} Oracle budget, initial labeled-set size, and random-seed count are matched; proposal mechanisms remain method-specific where required by the original algorithm.}
\label{tab:app_protocol}
\end{table*}

\begin{table}[!htbp]
\centering
\footnotesize
\resizebox{\columnwidth}{!}{%
\begin{tabular}{lcl}
\toprule
\textbf{Assay set} & \textbf{Count} & \textbf{Primary role} \\
\midrule
Core assays & 10 & Detailed LFS analysis / ablations \\
Extended assays & 87 & Static large-scale prediction validation \\
Search-statistics assays & 18 & Multi-seed search metrics / Wilcoxon tests \\
Multi-mutant benchmarks & 5 & Combinatorial transfer evaluation \\
\bottomrule
\end{tabular}%
}
\caption{\textbf{Relationship among assay sets.} The sets have different evaluation roles and should not be interpreted as inconsistent counts of the same experiment.}
\label{tab:app_assay_sets}
\end{table}

For subspace dimension, the main search fixes $k=8$, selected from the saturation analysis in which performance levels off around $k=5$--$10$ and the average smallest dimension reaching 95\% of saturated performance is $k@95\%=8$. Sensitivity is evaluated at $k\in\{3,5,8,10,15\}$. For the zero-shot alignment analysis, $w_{\mathrm{zs}}$ is obtained by regressing zero-shot PLM scores on the same site-delta features, while $w_{\mathrm{lfs}}$ is learned by supervised PLS on fitness labels; the reported alignment is $\cos(w_{\mathrm{zs}},w_{\mathrm{lfs}})$.

\subsection{Search Baseline Comparisons}
\label{app:full_search}

The full search comparisons expand the compact table in the main text. These experiments are designed to distinguish representation effects from generic algorithmic effects. Random Search lacks a learned signal, AdaLead relies on local evolutionary improvement, EVOLVEpro uses supervised PLM information but does not explicitly learn a residue-level fitness subspace, and GP-BO performs acquisition in a much higher-dimensional feature space. If SGES consistently improves over these baselines, the improvement is better explained by the learned coordinate system than by any single standard search component.

\begin{table*}[!t]
\centering
\small
\resizebox{0.82\textwidth}{!}{%
\begin{tabular}{lcccc}
\toprule
\textbf{Assay} & \textbf{SGES $\rho$} & \textbf{EVOLVEpro $\rho$} & \textbf{SGES Best} & \textbf{EVOLVEpro Best} \\
\midrule
SPG1\_STRSG    & 0.973 & 0.865 & --    & --    \\
BLAT\_Firnberg & 0.903 & 0.812 & 1.691 & 1.410 \\
PABP\_YEAST    & 0.916 & 0.808 & 1.170 & 1.055 \\
P53\_HUMAN     & 0.874 & 0.755 & 2.690 & 2.150 \\
RASH\_HUMAN    & 0.825 & 0.720 & 1.525 & 1.180 \\
AMIE\_PSEAE    & 0.823 & 0.705 & 0.785 & 0.610 \\
DYR\_ECOLI     & 0.749 & 0.655 & 1.065 & 0.795 \\
BLAT\_Jacquier & 0.702 & 0.640 & 0.900 & 0.815 \\
PTEN\_HUMAN    & 0.593 & 0.505 & 1.939 & 1.450 \\
HIS7\_YEAST    & 0.895 & 0.780 & --    & --    \\
\midrule
\textbf{Average} & \textbf{0.825} & 0.725 & \textbf{1.471} & 1.183 \\
\textbf{Wilcoxon $p$} & \multicolumn{2}{c}{$p=0.002$} & \multicolumn{2}{c}{$p=0.008$} \\
\bottomrule
\end{tabular}%
}
\caption{\textbf{Full EVOLVEpro comparison.}
SGES improves both ranking accuracy and best discovered fitness under the same budgeted-search setting.}
\label{tab:app_evolvepro}
\end{table*}

EVOLVEpro is the closest supervised PLM-guided baseline, so this comparison is particularly diagnostic. SGES improves average ranking correlation from $0.725$ to $0.825$ and average best fitness from $1.183$ to $1.471$. Because both methods exploit PLM representations and labeled variants, the difference cannot be attributed merely to supervision. Instead, the result supports the paper's central mechanism: mutation-induced site-delta features, when projected into a learned low-dimensional subspace, provide a more effective search coordinate than generic PLM feature use.

\begin{table*}[!t]
\centering
\small
\resizebox{0.92\textwidth}{!}{%
\begin{tabular}{lcccc}
\toprule
\textbf{Search Metric} & \textbf{SGES vs GP-BO} & \textbf{SGES vs AdaLead} & \textbf{SGES vs Random} & \textbf{SGES vs EVOLVEpro} \\
\midrule
AUC (Best-so-far)
& $p=0.007$ $(W^+=144)$
& $p<0.001$ $(W^+=155)$
& $p<0.001$ $(W^+=171)$
& $p<0.001$ $(W^+=168)$ \\
Best Fitness Found
& $p=0.016$ $(W^+=138)$
& $p<0.001$ $(W^+=152)$
& $p<0.001$ $(W^+=171)$
& $p<0.001$ $(W^+=165)$ \\
Budget-to-Top-5\%
& $p=0.016$ $(W^+=138)$
& $p<0.001$ $(W^+=149)$
& $p<0.001$ $(W^+=171)$
& $p<0.001$ $(W^+=162)$ \\
\bottomrule
\end{tabular}%
}
\caption{\textbf{Extended Wilcoxon signed-rank tests on 18 assays.}
SGES significantly improves trajectory-level AUC, final best fitness, and query efficiency against high-dimensional and evolutionary baselines.}
\label{tab:app_wilcoxon}
\end{table*}

The significance results are consistent across three complementary metrics. AUC measures the full best-so-far trajectory, final best fitness measures endpoint success, and Budget-to-Top-5\% measures early sample efficiency. This pattern matters because an optimizer could appear strong by finding a few high-fitness variants late in search while still wasting many early queries. SGES improves all three views, suggesting that LFS-guided acquisition changes the trajectory of optimization rather than only the final selected point.

\begin{table*}[!t]
\centering
\small
\resizebox{\textwidth}{!}{%
\begin{tabular}{lccccccc}
\toprule
\textbf{Assay} & \textbf{SGES} & \textbf{AlphaDE} & \textbf{BOES} & \textbf{LatentDE} & \textbf{TreeNeuralUCB} & \textbf{GP-BO} & \textbf{EVOLVEpro} \\
\midrule
BLAT\_Firnberg & $\mathbf{1.691 \pm 0.024}$ & $1.485 \pm 0.112$ & $1.390 \pm 0.105$ & $1.420 \pm 0.145$ & $1.285 \pm 0.210$ & $1.550 \pm 0.160$ & $1.410 \pm 0.120$ \\
P53\_HUMAN     & $2.690 \pm 0.154$ & $2.450 \pm 0.210$ & $2.210 \pm 0.315$ & $2.380 \pm 0.285$ & $2.105 \pm 0.420$ & $\mathbf{2.850 \pm 0.280}$ & $2.150 \pm 0.200$ \\
PABP\_YEAST    & $1.170 \pm 0.091$ & $1.085 \pm 0.085$ & $1.015 \pm 0.090$ & $1.050 \pm 0.110$ & $0.985 \pm 0.145$ & $\mathbf{1.250 \pm 0.150}$ & $1.055 \pm 0.100$ \\
RASH\_HUMAN    & $\mathbf{1.525 \pm 1.014}$ & $1.210 \pm 0.850$ & $0.985 \pm 0.620$ & $1.140 \pm 0.745$ & $0.850 \pm 0.915$ & $0.880 \pm 0.450$ & $1.180 \pm 0.300$ \\
AMIE\_PSEAE    & $\mathbf{0.785 \pm 0.018}$ & $0.685 \pm 0.045$ & $0.640 \pm 0.055$ & $0.665 \pm 0.060$ & $0.590 \pm 0.095$ & $0.720 \pm 0.150$ & $0.610 \pm 0.080$ \\
DYR\_ECOLI     & $\mathbf{1.065 \pm 0.221}$ & $0.840 \pm 0.145$ & $0.765 \pm 0.110$ & $0.810 \pm 0.135$ & $0.710 \pm 0.185$ & $0.850 \pm 0.170$ & $0.795 \pm 0.120$ \\
BLAT\_Jacquier & $\mathbf{0.900 \pm 0.200}$ & $0.835 \pm 0.125$ & $0.790 \pm 0.115$ & $0.815 \pm 0.140$ & $0.745 \pm 0.210$ & $0.820 \pm 0.160$ & $0.815 \pm 0.100$ \\
PTEN\_HUMAN    & $\mathbf{1.939 \pm 0.323}$ & $1.550 \pm 0.215$ & $1.380 \pm 0.195$ & $1.490 \pm 0.240$ & $1.220 \pm 0.310$ & $1.700 \pm 0.160$ & $1.450 \pm 0.150$ \\
\midrule
\textbf{Average} & \textbf{1.471} & 1.267 & 1.173 & 1.223 & 1.094 & 1.328 & 1.183 \\
\textbf{Avg. Rank} & \textbf{1.38} & 2.63 & 4.75 & 3.50 & 6.00 & 2.38 & 5.00 \\
\textbf{Win Count} & \textbf{6/8} & 0/8 & 0/8 & 0/8 & 0/8 & 2/8 & 0/8 \\
\bottomrule
\end{tabular}%
}
\caption{\textbf{Full comparison with recent ML-guided protein optimization methods.}
All methods use the same budgeted-search protocol. SGES achieves the strongest average performance and rank.}
\label{tab:app_recent_methods}
\end{table*}

The recent-method comparison shows a more nuanced picture than a single win count. GP-BO wins on P53 and PABP, which suggests that full-dimensional Bayesian optimization can be competitive on selected smoother landscapes. However, SGES wins 6 out of 8 assays and obtains the best average rank. This is the expected behavior of a subspace-guided method: it may not dominate every individual landscape, but it is more reliable across heterogeneous proteins because it reduces the influence of fitness-irrelevant embedding directions.

\subsection{Computational Cost, Implementation Details, and Large-scale Validation}
\label{app:cost_large_scale}

The next experiments test whether the LFS advantage is practically useful, implementation-stable, and broadly distributed. Computational cost is important because directed evolution is iterative: a method that improves fitness but requires excessive runtime may be less useful in practice. Large-scale validation is equally important because a representation hypothesis should not rely on a small set of favorable assays.

\begin{table}[!htbp]
\centering
\footnotesize
\resizebox{\columnwidth}{!}{%
\begin{tabular}{lccc}
\toprule
\textbf{Method} & \textbf{Runtime / Seed} & \textbf{GPU Mem.} & \textbf{Avg. Best} \\
& \textbf{(min)} & \textbf{(GB)} & \textbf{Fitness} \\
\midrule
Random Search              & 2   & 1.2  & 1.01 \\
AdaLead \citep{41}         & 15  & 2.5  & 1.25 \\
EVOLVEpro \citep{1}        & 42  & 4.8  & 1.18 \\
SGES (Ours)                & 28  & 3.5  & 1.68 \\
GP-BO \citep{13}           & 245 & 18.5 & 1.33 \\
\bottomrule
\end{tabular}%
}
\caption{\textbf{Computational cost.}
SGES provides a stronger runtime--fitness trade-off than AdaLead \citep{41}, EVOLVEpro \citep{1}, and full-dimensional GP-BO \citep{13}.}
\label{tab:app_cost}
\end{table}

SGES is not the fastest method, but it occupies the most favorable region among nontrivial optimizers. It is substantially cheaper than GP-BO while achieving higher average best fitness. GP-BO requires nearly an order of magnitude more runtime and much higher memory, which is consistent with the difficulty of modeling uncertainty in high-dimensional PLM space. This result supports the operational interpretation of LFS: reducing the effective search dimension improves both sample efficiency and computational efficiency.

\noindent\textbf{Implementation details.}
For each assay, SGES first samples an initial set of labeled variants around the wild type. Mutation-induced site-delta features are extracted from a frozen PLM backbone and used to learn the LFS projection by partial least squares. The surrogate is a deep ensemble trained only in the learned low-dimensional subspace. Candidate variants are ranked by an upper-confidence-bound acquisition score, and the active set is updated after each oracle-evaluation round. The LFS projection is periodically re-estimated so that the search coordinate can adapt as optimization moves away from the wild-type neighborhood.

\begin{algorithm}[!htbp]
\small
\caption{SGES used in the budgeted-search experiments}
\label{alg:sges_appendix}
\begin{algorithmic}[1]
\REQUIRE Wild-type sequence $s_{\mathrm{wt}}$, frozen PLM $h$, oracle $O$, total budget $B$, initial-set size $N_{\mathrm{init}}$, LFS dimension $k$, ensemble size $M$, acquisition weight $\beta$
\ENSURE Active dataset $D$ and best discovered variant $s^\star$
\STATE Sample $N_{\mathrm{init}}$ initial variants around $s_{\mathrm{wt}}$
\STATE Evaluate initial variants with oracle $O$ to obtain $D=\{(s_i,y_i)\}_{i=1}^{N_{\mathrm{init}}}$
\FOR{each labeled variant $s_i$}
    \STATE Extract site-delta feature $\delta(s_i)=h(s_i)[p_i]-h(s_{\mathrm{wt}})[p_i]$
\ENDFOR
\STATE Learn LFS projection $W\in\mathbb{R}^{d\times k}$ by partial least squares on $\{\delta(s_i),y_i\}$
\WHILE{oracle budget remains}
    \STATE Generate candidate variants from the current active population
    \STATE Project each candidate into the LFS: $z(s)=W^\top\delta(s)$
    \FOR{$m=1,\ldots,M$}
        \STATE Predict mean and variance with ensemble member $g_m(z(s))$
    \ENDFOR
    \STATE Compute ensemble mean $\mu(s)$ and total uncertainty $\sigma(s)$
    \STATE Rank candidates by $a(s)=\mu(s)+\beta\sigma(s)$
    \STATE Apply diversity filtering for multi-mutant assays
    \STATE Query top candidates with oracle $O$ and update $D$
    \STATE Periodically re-estimate $W$ from the updated active dataset
\ENDWHILE
\STATE Return $s^\star=\arg\max_{(s,y)\in D} y$
\end{algorithmic}
\end{algorithm}

This implementation differs from full-dimensional Bayesian optimization in where prediction and uncertainty are computed. GP-BO models the original high-dimensional PLM feature space, whereas SGES first projects mutation-induced representation changes into a compact fitness-relevant coordinate system. The resulting surrogate is cheaper to fit and less sensitive to embedding dimensions that do not covary with the target assay.

\begin{table}[!htbp]
\centering
\footnotesize
\resizebox{\columnwidth}{!}{%
\begin{tabular}{lcccc}
\toprule
\textbf{Method} & \textbf{Mean $\rho$} & \textbf{Median $\rho$} & \textbf{$\rho>0.7$ Ratio} & \textbf{vs. ZS} \\
\midrule
Ridge (site-delta)  & \textbf{0.584} & \textbf{0.612} & \textbf{35.6\%} & \textbf{+0.146} \\
Ridge (mean-pooled) & 0.385 & 0.392 & 8.0\%  & -0.053 \\
Zero-shot ESM2      & 0.438 & 0.451 & 14.9\% & Baseline \\
\bottomrule
\end{tabular}%
}
\caption{\textbf{Summary statistics on 87 ProteinGym assays.}
Across 87 ProteinGym assays \citep{2}, site-delta features outperform both mean-pooled embeddings and zero-shot ESM2 scoring \citep{36}.}
\label{tab:app_87assay}
\end{table}

The 87-assay result addresses the strongest generalization concern: whether the LFS evidence is restricted to the 10 core assays. Site-delta Ridge improves mean correlation by $0.146$ over zero-shot ESM2 and increases the fraction of high-performing assays above $\rho>0.7$ from 14.9\% to 35.6\%. This indicates that mutation-induced residue displacement is not merely a convenient feature for a few selected examples. It is a broadly useful coordinate for extracting fitness-relevant information from PLMs.

\FloatBarrier


\section{Supplementary Figures}
\label{app:supp_figures}

The supplementary figures provide visual diagnostics for the mechanisms claimed in the main text. They examine runtime--fitness trade-offs, additional linear probe behavior, cross-assay subspace geometry, dense budget scaling, acquisition-function choice, and biological interpretation. Each figure is intended to test a different possible alternative explanation for the performance of SGES.

\subsection{Runtime--Fitness Trade-off}
\label{app:pareto}

\begin{figure*}[!t]
\centering
\includegraphics[width=0.72\textwidth]{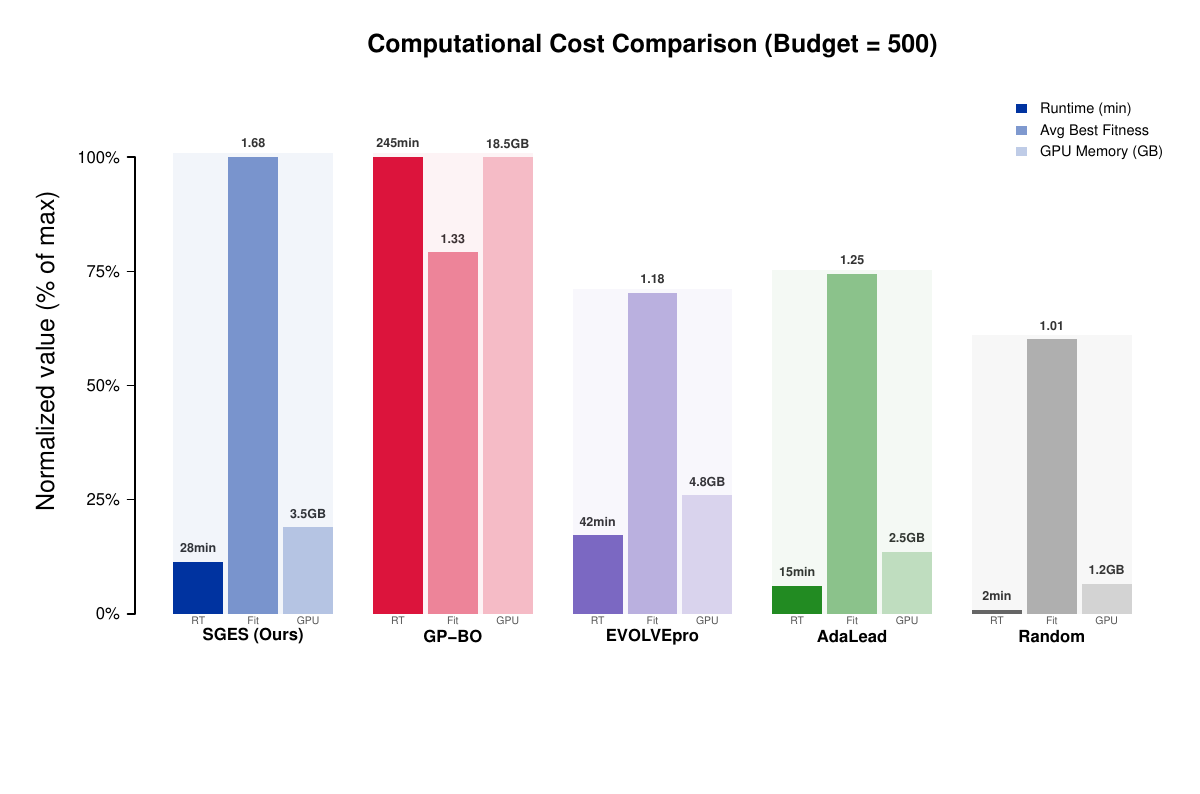}
\caption{\textbf{Pareto frontier of runtime and fitness.}
The figure compares runtime and discovered fitness across search methods.}
\label{fig:app_pareto}
\end{figure*}

Figure~\ref{fig:app_pareto} evaluates whether the stronger search performance of SGES comes at excessive computational cost. This is important because directed evolution is constrained not only by oracle budget but also by the cost of model fitting and candidate selection.

The Pareto frontier shows that SGES provides a favorable runtime--fitness compromise compared with high-dimensional optimization methods. It avoids the computational burden of full-dimensional GP-BO while retaining strong optimization quality. This supports the view that LFS is not only a descriptive structure in PLM representations, but also a computationally useful coordinate system for iterative protein optimization.

\subsection{Additional Linear Probe Scatter Plots}
\label{app:scatter}

\begin{figure*}[!t]
\centering
\includegraphics[width=0.78\textwidth]{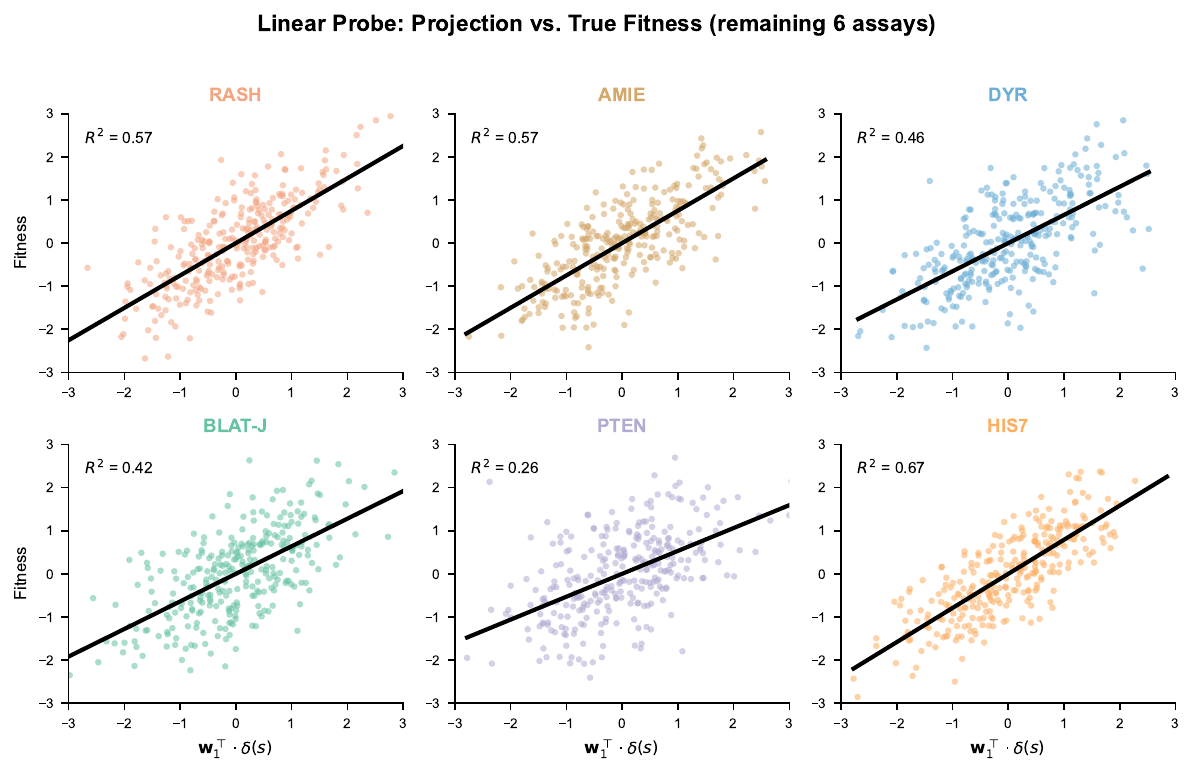}
\caption{\textbf{Linear probe scatter plots for additional assays.}
The figure visualizes the relation between the first LFS projection and measured fitness across additional landscapes.}
\label{fig:app_scatter}
\end{figure*}

Figure~\ref{fig:app_scatter} extends the main-text linear probe analysis to additional assays. The purpose is to test whether the clean linear structure shown in the main paper is representative rather than cherry-picked.

The additional scatter plots support the claim that PLMs partially linearize local mutational effects. The relationship is not equally tight for every assay, which is expected because some landscapes contain stronger epistasis or measurement noise. Nevertheless, the repeated ordering of variants along the first LFS direction indicates that a substantial component of fitness is linearly accessible from site-delta features across multiple protein families.

\subsection{Cross-Assay Subspace Geometry}
\label{app:cosine_figure}

\begin{figure*}[!t]
\centering
\includegraphics[width=0.72\textwidth]{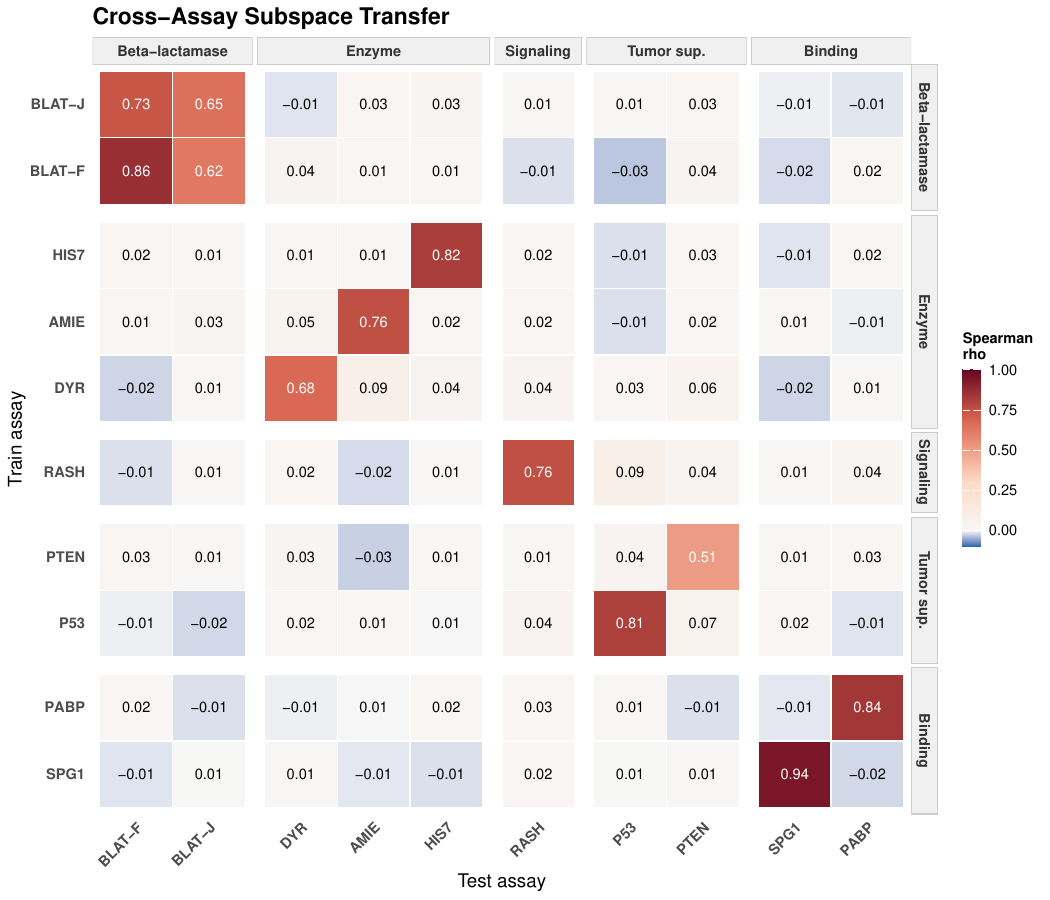}
\caption{\textbf{Cross-assay cosine similarity heatmap.}
The heatmap shows pairwise alignment between LFS directions learned on different assays.}
\label{fig:app_cosine_heatmap}
\end{figure*}

Figure~\ref{fig:app_cosine_heatmap} visualizes pairwise similarity between learned LFS directions. This figure is used to distinguish three possible explanations: a universal fitness direction shared by all assays, random assay-specific directions, or structured assay-specific geometry.

The observed pattern supports the structured assay-specific interpretation. Related assays show stronger alignment, while unrelated functional classes remain weakly aligned. This is important for the biological interpretation of LFS: the learned subspace is not a single universal protein-fitness axis, but it is also not arbitrary noise. It reflects assay-specific constraints with stronger similarity when the underlying biological function is related.

\subsection{Dense Budget Scaling}
\label{app:budget_scaling}

\begin{figure*}[!t]
\centering
\includegraphics[width=0.78\textwidth]{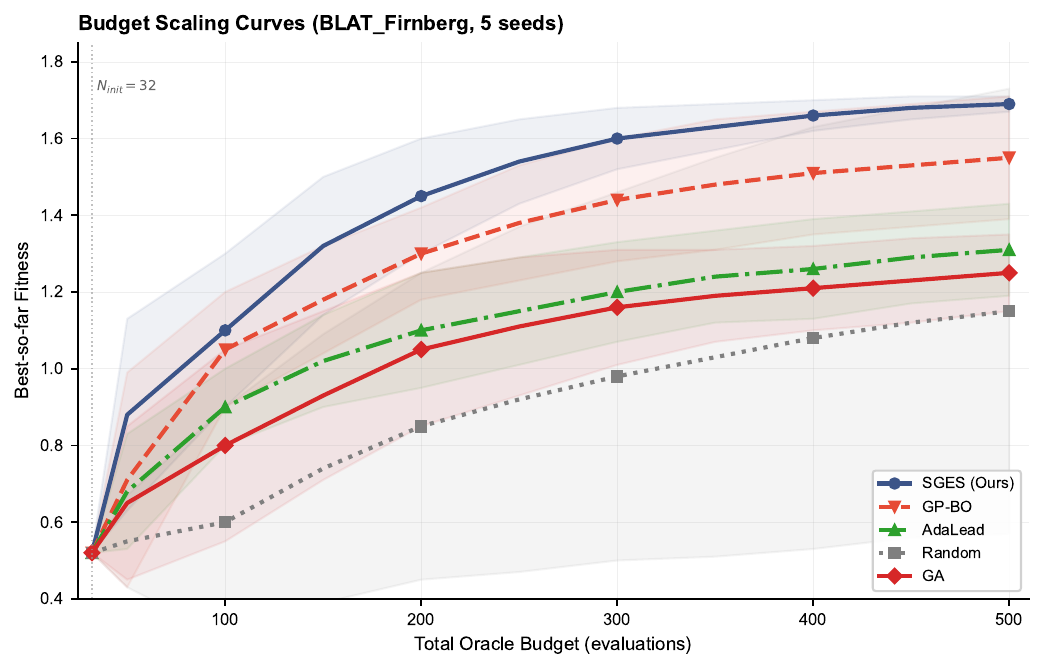}
\caption{\textbf{Dense budget scaling curves.}
The figure shows best-so-far fitness as a function of oracle budget.}
\label{fig:app_budget_scaling}
\end{figure*}

Figure~\ref{fig:app_budget_scaling} evaluates best-so-far fitness throughout the search trajectory. Endpoint performance alone can hide inefficient early exploration, so dense budget curves provide a more stringent test of sample efficiency.

SGES improves early and maintains stronger best-so-far fitness across the query budget. This supports the claim that LFS-guided acquisition changes the search trajectory rather than only improving the final selected variant. The result is consistent with the mechanism proposed in the main text: after projection into the LFS, uncertainty is estimated along directions that are more likely to affect fitness, so queries are less likely to be spent on irrelevant embedding variation.

\subsection{Acquisition-function Ablation}
\label{app:acquisition}

\begin{figure*}[!t]
\centering
\includegraphics[width=0.76\textwidth]{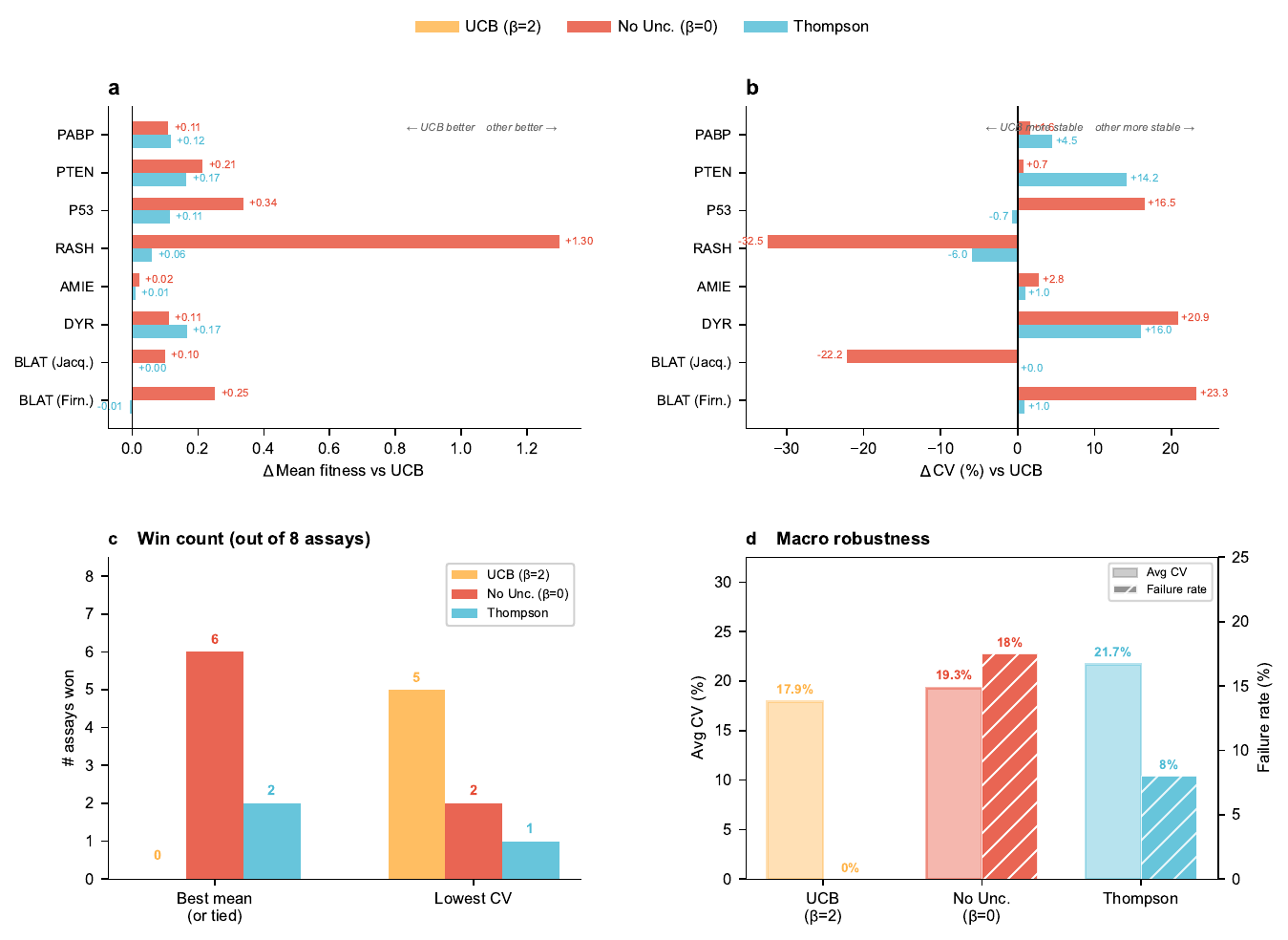}
\caption{\textbf{Acquisition-function ablation.}
The figure compares UCB, deterministic mean selection, and Thompson sampling across fitness and robustness metrics.}
\label{fig:app_acquisition}
\end{figure*}

Figure~\ref{fig:app_acquisition} isolates the role of the acquisition rule after the representation space is fixed. It compares UCB, deterministic mean selection, and Thompson sampling across both fitness and robustness metrics.

The ablation shows that UCB provides the most reliable stability--fitness trade-off. Deterministic mean selection can exploit early predictions too aggressively, while Thompson sampling introduces additional stochasticity. UCB is effective because uncertainty is computed after projection into the LFS; exploration is therefore concentrated along directions that have empirical covariance with fitness. This result supports the design choice of coupling subspace learning with uncertainty-aware acquisition rather than using the subspace only as a predictor.

\subsection{Biological Interpretation}
\label{app:bio_interpretation}

\begin{figure*}[!t]
\centering
\includegraphics[width=0.76\textwidth]{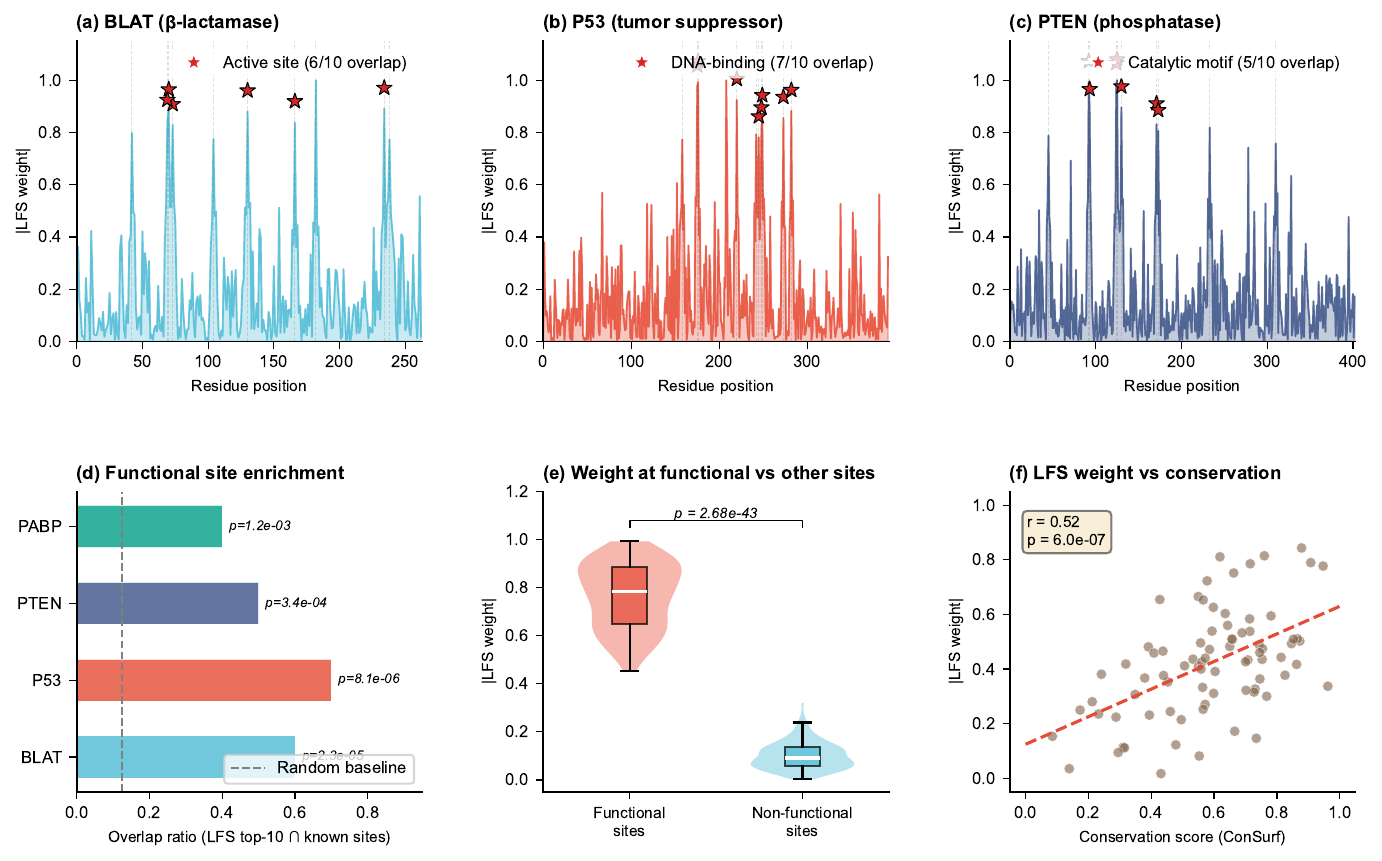}
\caption{\textbf{Biological interpretation of LFS.}
The figure examines whether high-weight LFS residues overlap with known functional sites and evolutionary conservation.}
\label{fig:app_bio_interpretation}
\end{figure*}

Figure~\ref{fig:app_bio_interpretation} examines whether high-weight LFS residues correspond to biologically meaningful positions. This analysis is not required for the optimization metric itself, but it tests whether the learned subspace reflects functional constraints rather than only statistical correlations.

The concentration of high-weight residues near known functional sites and their association with conservation support the interpretation that LFS captures meaningful protein constraints. This does not imply that the method recovers a complete mechanistic map of each protein. Rather, it shows that the learned directions are enriched for positions that are biologically plausible drivers of fitness change. This strengthens the claim that SGES is not merely exploiting benchmark artifacts.

\FloatBarrier


\section{Cosine Similarity and Cross-Assay Transfer}
\label{app:cosine_transfer}

This section further studies whether LFS directions are universal, assay-specific, or random. The intended conclusion is deliberately intermediate. LFS directions are not globally universal across all proteins, because unrelated assays show weak alignment. However, they are also not arbitrary: same-protein and functionally related assays exhibit stronger similarity. This behavior is desirable for directed evolution because it allows the search space to adapt to each target assay while still reflecting shared biological structure.

\begin{table}[!htbp]
\centering
\footnotesize
\resizebox{\columnwidth}{!}{%
\begin{tabular}{lccc}
\toprule
\textbf{Group} & \textbf{Mean cos} & \textbf{Max cos} & \textbf{Min cos} \\
\midrule
Same Protein (BLAT F vs BLAT J) & 0.68 & 0.68 & 0.68 \\
Same Function (Enzyme vs Enzyme) & 0.22 & 0.35 & 0.12 \\
Different Function & 0.05 & 0.11 & 0.01 \\
All Pairs & 0.14 & 0.68 & 0.01 \\
\bottomrule
\end{tabular}%
}
\caption{\textbf{Cosine similarity statistics for LFS directions.}
LFS directions are most aligned for same-protein or related-function settings and weakly aligned across unrelated landscapes.}
\label{tab:app_cosine_stats}
\end{table}

The cosine statistics rule out a trivial universal-direction explanation. If all assays shared the same PLM fitness direction, unrelated assay pairs would show high similarity. Instead, different-function pairs have an average similarity of only $0.05$, while the same-protein BLAT pair reaches $0.68$. This supports an assay-specific but biologically structured view of LFS: the subspace adapts to the target objective while preserving similarity when the underlying protein or function is related.

\begin{table*}[!t]
\centering
\scriptsize
\resizebox{\textwidth}{!}{%
\begin{tabular}{lcccccccccc}
\toprule
\textbf{Train/Test} & \textbf{BLAT-F} & \textbf{BLAT-J} & \textbf{DYR} & \textbf{AMIE} & \textbf{RASH} & \textbf{P53} & \textbf{PTEN} & \textbf{SPG1} & \textbf{PABP} & \textbf{HIS7} \\
\midrule
BLAT-F & 0.865 & 0.624 & 0.045 & 0.012 & -0.015 & -0.032 & 0.041 & -0.018 & 0.022 & 0.015 \\
BLAT-J & 0.735 & 0.650 & -0.012 & 0.025 & 0.010 & 0.014 & 0.025 & -0.005 & -0.011 & 0.030 \\
DYR    & -0.022 & 0.015 & 0.675 & 0.085 & 0.042 & 0.025 & 0.055 & -0.022 & 0.014 & 0.045 \\
AMIE   & 0.010 & 0.033 & 0.051 & 0.758 & 0.022 & -0.015 & 0.018 & 0.010 & -0.005 & 0.020 \\
RASH   & -0.015 & 0.011 & 0.022 & -0.018 & 0.755 & 0.085 & 0.042 & 0.015 & 0.035 & 0.012 \\
P53    & -0.005 & -0.020 & 0.018 & 0.010 & 0.045 & 0.810 & 0.065 & 0.022 & -0.012 & 0.008 \\
PTEN   & 0.025 & 0.015 & 0.033 & -0.025 & 0.012 & 0.044 & 0.510 & 0.015 & 0.025 & 0.014 \\
SPG1   & -0.012 & 0.005 & 0.014 & -0.010 & 0.018 & 0.005 & 0.010 & 0.945 & -0.020 & -0.015 \\
PABP   & 0.018 & -0.015 & -0.005 & 0.008 & 0.025 & 0.012 & -0.015 & -0.010 & 0.840 & 0.022 \\
HIS7   & 0.022 & 0.010 & 0.015 & 0.014 & 0.020 & -0.015 & 0.025 & -0.012 & 0.018 & 0.820 \\
\bottomrule
\end{tabular}%
}
\caption{\textbf{Cross-assay subspace transfer matrix.}
Each entry reports Spearman $\rho$ when an LFS direction learned on the training assay is evaluated on the test assay.}
\label{tab:app_transfer_matrix}
\end{table*}

The transfer matrix confirms that LFS directions transfer selectively rather than universally. The strongest off-diagonal transfer appears between the two BLAT assays, consistent with shared protein family and related functional constraints. Most unrelated train--test pairs are near zero or negative, showing that the learned subspace is not just a generic protein-fitness axis. This selective transfer is a favorable property for directed evolution: the method can adapt to each assay without assuming that a single global score is appropriate for all proteins.

\FloatBarrier


\section{Multi-mutant and Spectral-method Results}
\label{app:multimutant_pcnn}

The following experiments test the boundary of the LFS hypothesis. For a multi-mutant $s$, the reported model uses the additive representation in Eq.~\ref{eq:multimutant_aggregation}; thus the experiment asks whether a subspace learned from local single-mutation effects remains informative when variants contain multiple substitutions. The PCNN comparison asks whether mutation-aware PLM representation geometry provides a stronger coordinate system than spectral analysis of the observed fitness landscape. Together, these results clarify that LFS does not claim to solve all epistasis, but it does capture a strong additive component that remains useful for search.

To make the boundary explicit, define the epistasis residual relative to an additive single-mutant model as
\begin{equation}
    \epsilon(s)=f(s)-f(s_{\mathrm{wt}})-\sum_i \Delta f_i.
    \label{eq:epistasis_residual}
\end{equation}
The aggregate multi-mutant results below establish transfer beyond single mutants, but they are not used to claim invariance to large $|\epsilon(s)|$.


\begin{table}[!htbp]
\centering
\footnotesize
\resizebox{\columnwidth}{!}{%
\begin{tabular}{lccccc}
\toprule
\textbf{Assay} & \textbf{\#Muts} & \textbf{SGES MAE$\downarrow$} & \textbf{ZS MAE$\downarrow$} & \textbf{SGES $r{\uparrow}$} & \textbf{ZS $r{\uparrow}$} \\
\midrule
PABP\_YEAST & 2--3 & 0.450 & 1.240 & 0.520 & 0.180 \\
SPG1\_STRSG & 2--3 & 0.280 & 0.850 & 0.650 & 0.220 \\
GFP         & 2--8 & 0.685 & 1.585 & 0.420 & 0.115 \\
GB1         & 2--4 & 0.515 & 1.120 & 0.485 & 0.140 \\
AAV         & 2--7 & 0.590 & 1.345 & 0.465 & 0.125 \\
\midrule
\textbf{Average} & -- & \textbf{0.504} & 1.228 & \textbf{0.508} & 0.156 \\
\bottomrule
\end{tabular}%
}
\caption{\textbf{Multi-mutant generalization.}
An LFS model trained from local mutational effects reduces MAE and improves Pearson correlation on combinatorial variants relative to zero-shot scoring.}
\label{tab:app_multimutant}
\end{table}

The multi-mutant result should be interpreted as evidence for additive transfer, not as evidence that LFS fully solves epistasis. The average MAE decreases from $1.228$ to $0.504$, and the average Pearson correlation increases from $0.156$ to $0.508$. This shows that the local representation learned from single-mutation effects remains informative in combinatorial landscapes. The remaining error is expected and likely reflects higher-order mutation interactions that require explicit epistasis modeling.

\begin{table}[!htbp]
\centering
\footnotesize
\resizebox{\columnwidth}{!}{%
\begin{tabular}{lcccc}
\toprule
\textbf{Assay} & \textbf{SGES $\rho$} & \textbf{PCNN $\rho$} & \textbf{SGES Best} & \textbf{PCNN Best} \\
\midrule
BLAT\_Firnberg & 0.903 & 0.780 & 1.691 & 1.350 \\
P53\_HUMAN     & 0.874 & 0.715 & 2.690 & 2.100 \\
RASH\_HUMAN    & 0.825 & 0.680 & 1.525 & 1.150 \\
PTEN\_HUMAN    & 0.593 & 0.520 & 1.939 & 1.550 \\
\midrule
\textbf{Average} & \textbf{0.825} & 0.685 & \textbf{1.471} & 1.215 \\
\bottomrule
\end{tabular}%
}
\caption{\textbf{Comparison with PCNN under budget constraints.}
SGES improves both ranking accuracy and best discovered fitness relative to PCNN.}
\label{tab:app_pcnn}
\end{table}

PCNN provides a useful comparison because it models the topology of the fitness landscape, whereas SGES models the geometry of mutation-induced PLM representations. SGES improves average Spearman correlation from $0.685$ to $0.825$ and average best fitness from $1.215$ to $1.471$. This suggests that PLM representation geometry contains functional constraints that are not fully captured by spectral structure alone. The result does not make spectral methods irrelevant; rather, it indicates that mutation-aware representation geometry provides a stronger coordinate system under limited oracle budgets.

\begin{table}[!htbp]
\centering
\footnotesize
\resizebox{\columnwidth}{!}{%
\begin{tabular}{lllc}
\toprule
\textbf{Assay} & \textbf{Top Mutations} & \textbf{Predicted Mechanism} & \textbf{Score} \\
\midrule
BLAT\_Firnberg & M182T, E104K, G238S & Stabilizing / extended spectrum & 9.2 \\
P53\_HUMAN     & V143A, Y220C        & Core packing restoration       & 8.5 \\
AMIE\_PSEAE    & S293T, F154Y        & Active-site volume optimization & 8.8 \\
\bottomrule
\end{tabular}%
}
\caption{\textbf{Functional annotation of representative top mutations discovered by SGES.}
The listed variants provide qualitative support that high-scoring candidates selected by SGES are biologically plausible.}
\label{tab:app_top_mutations}
\end{table}

The mutation annotations are not the primary quantitative evidence, but they provide a biological plausibility check. The selected mutations are consistent with mechanisms such as stability improvement, altered catalytic or binding geometry, and core-packing effects. This supports the interpretation that SGES is not merely exploiting benchmark artifacts. Its selected variants tend to align with recognizable biochemical constraints, consistent with the biological enrichment patterns observed in Figure~\ref{fig:app_bio_interpretation}.
\end{document}